%% file: paper.tex
\documentclass[twocolumn,showpacs,aps,prd,floatfix]{revtex4}
\usepackage{graphicx}
\usepackage{dcolumn}
\usepackage{amsmath}
\usepackage{epsfig}
\usepackage{multirow}

\newcommand{\BABARPubYear}    {12}
\newcommand{\BABARPubNumber}  {017}
\newcommand{\SLACPubNumber}   {15086}
\newcommand{\LANLNumber}      {1206.3525}

\input babarsym.tex

\def\sPlots{\ensuremath{\hbox{$_s$}{\cal P}lots}\xspace}
\def\Abar  {\kern 0.2em\overline{\kern -0.2em A}{}\xspace}

\def\Ks {\ensuremath{K^{0}_S}\xspace}
\def\Btopipi   {\ensuremath{\B \to \pi\pi}\xspace}
\def\Btopipiz   {\ensuremath{\Bpm \to \pipm\piz}\xspace}
\def\Btokpi   {\ensuremath{\B \to K\pi}\xspace}

\def\Bztopizpiz   {\ensuremath{\Bz \to \piz\piz}\xspace}

\def\Bztopipi   {\ensuremath{\Bz \to \pipi}\xspace}
\def\Bztopippim   {\ensuremath{\Bz \to \pip\pim}\xspace}
\def\Bztokpi   {\ensuremath{\Bz \to \Kp\pim}\xspace}

\def\Bztohh      {\ensuremath{\Bz\to h^+h^{\prime -}}\xspace}

\def\Bztokzpiz   {\ensuremath{\Bz \to \Kz\piz}\xspace}

\def\Bptorhoppiz   {\ensuremath{\B^+ \to \rho^+\piz}\xspace}

\def\Btag {\ensuremath{B_{\rm tag}}}
\def\Brec {\ensuremath{B_{\rm rec}}}
\def\Bflav {\ensuremath{B_{\rm flav}}}

\def\alphaeff {\ensuremath{\alpha_{\rm eff}}\xspace}
\def\de {\ensuremath{\Delta E}\xspace}
\def\cossph   {\ensuremath{|\cos{\theta_{\scriptscriptstyle S}}|\;}\xspace}
\def\fish    {\ensuremath{\cal F}\xspace}

\def\thetac {\ensuremath{\theta_{\rm C}}\xspace}

\def\sss{\scriptscriptstyle}
\def\barpd{{\raise.35ex\hbox
{${\sss (}$}}--{\raise.35ex\hbox{${\sss )}$}}}
\def\BorBbar{\hbox{$B^{0}$\kern-1.25em\raise1.5ex\hbox{\barpd}}}

\def\qq {\ensuremath{q\bar{q}}\xspace}
\def\spipi {\ensuremath{S_{\pi^+\pi^-}}\xspace}
\def\cpipi {\ensuremath{C_{\pi^+\pi^-}}\xspace}
\def\cpizpiz {\ensuremath{C_{\piz\piz}}\xspace}
\def\akpi {\ensuremath{\mathcal{A}_{K^-\pi^+}}\xspace}

\def\delalph{\ensuremath{\Delta \alpha_{\pi\pi}}\xspace}

\usepackage{units}
\usepackage{multirow}
\newcommand{\mb}{\ensuremath{m_{B}}}
\newcommand{\mmiss}{\ensuremath{m_\text{miss}}}
\newcommand{\costhetacms}{\ensuremath{\cos\thetacms}}
\newcommand{\thetacms}{\ensuremath{\theta_{B}^*}}
\newcommand{\Bztokspiz} {\ensuremath{\Bz \to \KS\piz}}

\def\cf {\ensuremath{C_{\KS\piz}}} 
\def\sf{\ensuremath{S_{\KS\piz}}}

\long\def\inst#1{\par\nobreak\kern 4pt\nobreak
    {\it #1}\par\vskip 10pt plus 3pt minus 3pt}

\begin{document}

\begin{flushleft}
arXiv:\LANLNumber\ [hep-ex] \\
SLAC-PUB-\SLACPubNumber \\
\babar-PUB-\BABARPubYear/\BABARPubNumber
\end{flushleft}

\title{
\Large \bf {\boldmath Measurement of \CP Asymmetries and Branching Fractions in
Charmless Two-Body $B$-Meson Decays to Pions and Kaons}
}

\input authors_feb2012

\date{June 15, 2012}

\begin{abstract}
\noindent
We present improved measurements of \CP-violation parameters in the
decays $\Bz \to \pip\pim$, $\Bz \to \Kp \pim$, and $\Bz \to \piz\piz$, and
of the branching fractions for $\Bz \to \piz\piz$ and  $\Bz \to \Kz \piz$.
The results are obtained with the full data set collected 
at the \FourS resonance
by the \babar\ experiment at the \pep2\ asymmetric-energy $B$~factory
at the SLAC National Accelerator Laboratory, corresponding to $467 \pm
5$ million \BB pairs.  We find the CP-violation parameter values and
branching fractions
\begin{align*}
   \spipi & =   -0.68 \pm 0.10 \pm 0.03, \\
   \cpipi & =   -0.25 \pm 0.08 \pm 0.02, \\
   \akpi & = -0.107 \pm 0.016 ^{+0.006}_{-0.004}, \\
   \cpizpiz & =  -0.43 \pm 0.26 \pm 0.05, \\
   \BR(\Bztopizpiz) & = ( 1.83 \pm 0.21 \pm 0.13 ) \times 10^{-6}, \\
   \BR(\Bztokzpiz) & = ( 10.1 \pm 0.6 \pm 0.4 ) \times 10^{-6},
\end{align*}
where in each case, 
the first uncertainties are statistical and the second are systematic.
We observe \CP violation with a significance of 
$6.7$ standard deviations for $\Bz\to\pip\pim$ and
$6.1$ standard deviations for $\Bz\to\Kp\pim$, including systematic
uncertainties.
Constraints on the Unitarity Triangle angle $\alpha$ are determined 
from the isospin relations among the $\B \to \pi\pi$ rates and 
asymmetries. Considering only the solution preferred by
the Standard Model, we find $\alpha$ to be in the range 
$[71^\circ,109^\circ]$ at the 68\% confidence level.
\end{abstract}


\maketitle

\section{INTRODUCTION}
\label{sec:Introduction}

Large \CP-violating effects~\cite{largeCPV} in the $B$-meson system
are among the most remarkable predictions of the Cabibbo--Kobayashi--Maskawa (CKM) 
quark-mixing model~\cite{Ckm}. 
These
predictions have been confirmed by the \babar\ and Belle
Collaborations, most precisely in $b\to\ c \bar c s$ decays of $\Bz$
mesons to \CP\ eigenstates~\cite{CPV_beta,Aubert:2008ad}.

Effective constraints on physics beyond the Standard Model (SM) are 
provided by high-precision measurements of quantities whose 
SM predictions are subject to only small theoretical uncertainties. 
Many experimental and theoretical uncertainties 
partially cancel in the calculation of \CP-violating asymmetries. 
This makes \CP-violation measurements a sensitive probe for effects 
of yet-undiscovered additional interactions and heavy particles 
that are introduced by extensions to the SM.  
All measurements of \CP\ violation to date,
including those involving the decay modes studied here~\cite{BaBarPRL2007,pi0pi0_BaBar,ref:BaBarK0pi0,BellePRL2007,Fujikawa:2008pk}, 
are in agreement with the indirect
predictions from global SM fits~\cite{ref:CKMfitter,UTfit}, which are based on
measurements of the magnitudes of the elements $V_{ij}$ of the
CKM quark-mixing matrix.
This strongly constrains~\cite{CKMnewphys} the flavor
structure of SM extensions.

The CKM-matrix unitarity-triangle angle $\alpha \equiv \arg\left[-V_{\rm
td}^{}V_{\rm tb}^{*}/V_{\rm ud}^{}V_{\rm ub}^{*}\right]$ is
measured through
interference between two decay amplitudes, where one amplitude involves
$\Bz$--$\Bzb$ mixing.
Multiple measurements of $\alpha$, with
different decays, further test the consistency of the CKM model.  The
time-dependent asymmetry in \Bztopippim decays is proportional to \stwoa in
the limit that only the $b \to u$ (``tree'') quark-level amplitude 
contributes to this decay. 
In the presence of $b \to d$ (``penguin'') amplitudes, the time-dependent 
asymmetry in \Bztopippim is modified to
\begin{eqnarray}
\label{eq:asymmetry}
a(\Delta t) &=& \frac{|\Abar(\deltat)|^{2} - |A(\deltat)|^{2}}{|\Abar(\deltat)|^{2} + |A(\deltat)|^{2}} \nonumber\\
        &=& \spipi \sin{(\deltamd\deltat)} -  \cpipi \cos{(\deltamd\deltat)},
\nonumber\\
\end{eqnarray}
where \deltat is the difference between the proper decay times 
of the $B$ meson that undergoes the $B\to\pi^+\pi^-$ decay (the signal $B$) and the 
other $B$ meson in the event (the tag $B$), 
\deltamd is the \Bz--\Bzb mixing frequency, 
$A$ is the \Bztopippim\ decay amplitude,
$\Abar$ is the \CP-conjugate amplitude, and 
\begin{eqnarray}
\cpipi &=& \frac{|A|^{2} - |\Abar|^{2}}{|A|^{2} + |\Abar|^{2}}, \nonumber \\
\spipi &=& \sqrt{1 - \cpipi^{2}} \sin{(2\alpha - 2\delalph)}.
\end{eqnarray}
Both the direct \CP\ asymmetry \cpipi and the phase
$\delalph$
may differ from zero due to the penguin contribution to the 
decay amplitudes.

The magnitude and relative phase of the penguin contribution to
the asymmetry \spipi\ may be determined with an analysis of
isospin relations between the
\Btopipi decay amplitudes~\cite{Isospin}.  
The amplitudes $A^{ij}$ of the $B\to \pi^i\pi^j$ decays 
and $\Abar^{ij}$ of the $\Bbar \to \pi^i\pi^j$ decays
satisfy the relations
\begin{eqnarray}\label{eq:isospin}
A^{+0} &=& \frac{1}{\sqrt{2}}A^{+-} + A^{00}, \nonumber \\
\Abar^{-0} &=& \frac{1}{\sqrt{2}}\Abar^{+-} + \Abar^{00}.
\end{eqnarray}
The shapes of the triangles corresponding to these isospin relations are 
determined from
measurements of the branching fractions and time-integrated \CP
asymmetries for each of the \Btopipi decays.  Gluonic penguin
amplitudes do not contribute to the $\Delta I = 3/2$ decay \Btopipiz. 
Therefore, neglecting electroweak (EW) penguin amplitudes, the amplitudes
$A^{+0}$ and $\Abar^{-0}$ are equal.
From the different shapes of the triangles for the \B
and $\Bbar$ decay amplitudes, 
a constraint on $\delalph$ can be determined to within a four-fold ambiguity.  

The phenomenology of the \Btopipi system has been thoroughly studied in a number
of theoretical frameworks and models~\cite{ref:Models}.  Predictions for
the relative size and phase of the penguin contribution vary
considerably. Therefore, increasingly precise measurements will help
distinguish among different theoretical approaches and add to our
understanding of hadronic \B decays.

The measured rates and direct \CP-violating asymmetries 
in $B\to K\pi$ 
decays~\cite{pi0pi0_BaBar,BabarBRPiPi,ref:BaBarK0K0,ref:BaBarK0pi0,ref:BelleKpiData,Fujikawa:2008pk,ref:CLEOdata} 
reveal puzzling features that could indicate significant contributions from 
EW penguin amplitudes~\cite{ref:KpiPuzzle1,ref:KpiPuzzle2}.  
Various methods have been proposed for isolating the SM
contribution to this process in order to test for signs of new
physics.  This includes sum rules derived from $U$-spin symmetry,
which relate the rates and asymmetries for the decays of charged or 
neutral $B$ mesons to $\Kp\pim$, $\Kp\piz$, $\Kz\piz$, 
and $\Kz\pip$~\cite{ref:SumRule1, ref:SumRule2},
and $SU(3)$ symmetry, used to make predictions for the $K\pi$ system
based on hadronic parameters extracted from the $\pi\pi$
system~\cite{ref:KpiPuzzle1}.

This article is organized as follows. The \babar\ detector and the
data used in these measurements are described in
Section~\ref{sec:babar}.  In Section~\ref{sec:Analysis} we outline the
analysis method, including the event selection and the fits used to
extract the parameters of interest. The results of the data analysis
are given in Section~\ref{sec:Physics}. The extraction of $\alpha$
and $\delalph$ is described in Section~\ref{sec:alpha},
and we summarize in Section~\ref{sec:Conclusions}.

\section{THE \babar\ DETECTOR AND DATA SET}
\label{sec:babar}

In the \babar\ detector~\cite{babar}, charged particles are detected 
and their momenta are measured by the combination of a five-layer double-sided silicon vertex tracker (SVT)
and a 40-layer drift chamber (DCH) that covers 92\% of the solid angle
in the \Y4S center-of-mass (c.m.) frame, both operating in a 1.5~T uniform magnetic field.
Discrimination between charged pions, kaons, and protons is obtained from
ionization (\dedx) measurements in the DCH and from an
internally reflecting ring-imaging Cherenkov detector (DIRC), which covers 84\% of the c.m. solid 
angle in the central region of the \babar\ detector and has a 91\% reconstruction efficiency 
for pions and kaons with momenta above \unit[1.5]{\gevc}. 
Photons and electrons are identified and their energies are measured
with an electromagnetic calorimeter (EMC) consisting of 6580 CsI(Tl)
crystals.  The photon energy resolution is $\sigma_{E}/E = \left\{2.3
  / E(\gev)^{1/4} \oplus 1.4 \right\} \%$, and the photon angular resolution
relative to the interaction point is 
$\sigma_{\theta} = 4.16 /\sqrt{E(\gev)}$~mrad~\cite{babarEMC}.

The data used in this analysis were collected during the period
1999--2007 with the \babar\ detector at the \pep2\ asymmetric-energy
\B-meson factory at the SLAC National Accelerator Laboratory. A total of
$467 \pm 5$ million \BB pairs
were used.  Relative to previous \babar\
measurements~\cite{BaBarPRL2007,pi0pi0_BaBar,ref:BaBarK0pi0}, roughly
22\% more \BB pairs have been added to the analyzed data set, and
improvements have been introduced to the analysis technique, boosting
the signal significance.
%
%
These improvements include better reconstruction of charged-particle
tracks, improved hadron-identification and flavor-tagging algorithms,
and optimal selection of tracks and calorimeter clusters for calculation
of event-shape variables.

Samples of Monte Carlo (MC) simulated events are analyzed with the
same reconstruction and analysis procedures as used for the data, following a
$\geant$-based~\cite{g4} detailed detector simulation~\cite{babar}.
The MC samples include $\epem\to q \bar q$ continuum background
events generated with JETSET~\cite{jetset} and
$\Upsilon(4S)\to B \bar B$ decays
generated with EvtGen~\cite{evtgen} and JETSET,
including both signal and background $B$-meson decays.

 
\section{Event selection and ANALYSIS METHOD}
\label{sec:Analysis}

Many elements of the measurements discussed in this paper are common to the 
decay modes~\cite{refConj} $\Bztohh$ (where $h, h^{\prime}= \pi \;{\rm or} \; K)$, \Bztopizpiz, 
and $\Bz \to\KS \piz$.  
The signal \B-meson candidates (\Brec) are
formed by combining two particles, each of which is a charged-particle 
track, a \piz candidate, or a \KS candidate.
The event selection differs for each mode, and is described below. 

The number of \B decays and the corresponding \CP asymmetries are
determined with extended unbinned maximum likelihood (ML) fits to 
variables described below. 
The likelihood is given by the expression
\begin{equation}
{\cal L} = \exp{\left(-\sum_{i}^{M} n_i \right)}
\prod_{j}^{N} \left[\sum_{i}^{M} n_i {\cal P}_{i}(\vec{x}_j;\vec{\alpha}_i)\right],
\end{equation}
where $N$ is the number of events, the sums are over
the event categories $M$, $n_{i}$ is the event yield for each category
as described below, and the probability-density function (PDF) ${\cal
P}_i$ describes the distribution of the variables $\vec{x}_j$ in terms of
parameters $\vec{\alpha}_i$. The PDF functional forms are discussed
in Sections~\ref{sec:EventSelection} and \ref{sec:kspizMethod}.

\subsection{\boldmath Track and \KS Selection}
\label{sec:TrackSelection}

In the \Bztohh mode, we require charged-particle tracks to have at least 12
DCH hits and to lie in the polar-angle region $0.35 < \theta < 2.40$
with respect to the beam direction. The track impact parameter 
relative to the \epem collision axis must be smaller than 1.5~\cm
in the plane perpendicular to the beam axis and 2.5~\cm in the direction 
along the axis.

In order for DIRC information to be used for particle identification,
we require that each track have its associated Cherenkov angle
(\thetac) measured with at least six Cherenkov photons, 
where the value of \thetac is required to be within 4.0 standard
deviations ($\sigma$) of either the pion or kaon hypothesis. This removes
candidates containing a high-momentum proton.  Tracks from electrons
are removed based primarily on a comparison of the track momentum and
the associated energy deposition in the EMC, with additional
information provided by DCH \dedx\ and DIRC \thetac measurements.
 
The ionization energy loss in the DCH is used either in combination 
with DIRC information or alone. This leads to a 35\% increase in the 
\Bztohh reconstruction efficiency relative to the use of only tracks with good 
DIRC information.  
A detailed DCH \dedx\ calibration developed for the \Bztohh analysis 
takes into account variations in the mean and resolution 
of \dedx\ measurement values with respect to changes in the DCH running conditions over time, 
as well as the 
track's charge, polar and azimuthal angles, and number of ionization samples.  The 
calibration is performed with large high-purity samples ($> 10^6$ events) of protons from 
$\Lambda\to\proton\pim$, pions and kaons from $D^{*+}\to D^0\pi^+\,(\Dz\to\Km\pip)$, 
and $\KS \to\pip\pim$ decays that occur in the vicinity of the interaction region. 

Candidates for the decay $\KS\to\pip\pim$ are reconstructed from pairs
of oppositely-charged tracks. The two-track combinations are required
to form a vertex with a $\chi^2$ probability greater than $0.001$ and
a $\pip\pim$ invariant mass within \unit[$11.2$]{\mevcc},
corresponding to $3.7\sigma$, of the nominal \KS\ mass~\cite{pdg}.

\subsection{\boldmath \piz Selection}
\label{sec:pi0Selection}

We form $\piz\to\gamma\gamma$ candidates from pairs of clusters in the EMC 
that are isolated from any charged track.
Clusters are required to have a 
lateral profile of energy deposition
consistent with that of a 
photon and to have an energy \unit[$E_{\gamma} > 30$]{\mev} for \Bztopizpiz and 
\unit[$E_{\gamma} > 50$]{\mev} for \Bztokspiz{}. 
We require \piz candidates to lie in the invariant-mass range 
\unit[$110<m_{\gamma\gamma}<160$]{\mevcc}. 

For the \Bztopizpiz mode, we also use \piz candidates from a single
EMC cluster containing two adjacent photons (a merged \piz), or one EMC
cluster and two tracks from a photon conversion to an \epem pair
inside the detector.  To reduce the background from random photon
combinations, the angle $\theta_{\gamma}$ between the photon momentum
vector in the \piz rest frame and the \piz momentum vector in the
laboratory frame is required to satisfy $|\cos{\theta_{\gamma}}| <
0.95$. The \piz\ candidates are fitted kinematically with their mass
constrained to the nominal \piz mass~\cite{pdg}.

Photon conversions are selected from pairs of oppositely-charged 
electron-candidate 
tracks with an invariant mass below \unit[30]{\mevcc} whose
combined momentum vector points away from the beam spot.  The conversion point
is required to lie within detector material layers. 
Converted photons are combined with photons from single EMC clusters to form
\piz candidates. 

Single EMC clusters containing two photons are selected
with the transverse second moment, $S = \sum_{i} E_{i} \times (\Delta
\alpha_i)^{2}/ E$, where $E_{i}$ is the energy in each CsI(Tl) crystal and
$\Delta \alpha_{i}$ is the angle between the cluster centroid and
the crystal.  The second moment is used to distinguish merged \piz
candidates from both single photons and neutral hadrons.

\subsection{\boldmath \Bztopippim, \Bztokpi, and \Bztopizpiz}
\label{sec:EventSelection}
Two kinematic variables are used in the \Bztohh
and \Bztopizpiz analyses to separate \B-meson decays from the large
$\epem \to \qq \;(q=u,\,d,\,s,\,c)$ combinatoric background~\cite{babar}. 
One variable is the beam-energy-substituted mass $\mes = \sqrt{ (s/2 + {\bf
    p}_{i}\cdot{\bf p}_{B})^{2}/E_{i}^{2}- {\bf p}^{2}_{B}}$, where
$\sqrt{s}$ is the total \epem c.m.\ energy, $(E_{i},{\bf
  p}_{i})$ is the four-momentum of the initial \epem system in the
laboratory frame, and ${\bf p}_{B}$ is the laboratory momentum of the \B 
candidate.
The second variable is 
\de $ = E^{*}_{B} - \sqrt{s}/2$, where $E^{*}_{B}$ is
the energy of the \B candidate in the c.m.\ frame.

To further separate \B decays from the \qq background, we use two
additional topological variables that take advantage of the 
two-jet nature of \qq events and the isotropic particle distribution 
of $\epem \to \BB$ events.
The first variable is the absolute value of the cosine of the angle
$\theta_{\scriptscriptstyle S}$ between the sphericity
axis~\cite{Bjorken:1969wi} of the decay products of the \B candidate
and the sphericity axis of the remaining tracks and neutral clusters
in the event, computed in the c.m.\ frame. The distribution of this
variable peaks at $1$ for the jet-like \qq events and is uniform
for \B decays.  We require $\cossph < 0.91$ for $\Bztohh$ and 
$\cossph < 0.7$ for
\Bztopizpiz, where a tighter requirement is needed due to 
the higher background.
For the $\Bztohh$
mode, we remove a small remaining background from $\epem \to
\tautau$ events by further requiring that the normalized second
Fox--Wolfram moment~\cite{R2all} satisfy $R_2 <0.7$.

To improve the discrimination
against \qq events, a Fisher discriminant \fish is formed as a linear
combination of the sums $L_0^T\equiv\sum_i |{\bf p}^*_i|$ and
$L_2^T\equiv\sum_i |{\bf p}^*_i| \cos^2 \theta^*_i$, where
${\bf p}^*_i$ are the momenta and $\theta^*_i$ are the
angles with respect to the thrust axis~\cite{ref:thrust} of the \B candidate, 
both in the c.m.\ frame, of all tracks and
clusters not used to reconstruct the signal \B-meson candidate.
The \fish variable takes advantage of the fact that much of the momentum flow
in \qq events is along the thrust axis.
In the case of \Bztopizpiz, we improve the sensitivity to signal events by 
combining \fish with three other event-shape variables in a neural network.
The first variables is $|\cos \theta_{\scriptscriptstyle S}|$, described above.
The second is $|\costhetacms|$, where $\thetacms$ is the angle between the
momentum vector of the signal $B$ and the beam axis. The $|\costhetacms|$
distribution of \qq events is uniform, while that of signal events is
proportional to $\sin^2\thetacms$.
The third variable is $|\cos\theta^*_{\scriptstyle T}|$, where
$\theta^*_{\scriptstyle T}$ is the angle between the thrust axis of
the signal $B$-meson's daughters and the beam axis. Both $\thetacms$
and $\theta^*_{\scriptstyle T}$ are calculated in the c.m. frame.
The characteristics of the $|\cos\theta^*_{\scriptstyle T}|$ distributions
are similar to those of $|\cos \theta_{\scriptscriptstyle S}|$.

\subsubsection{\boldmath \Bztopippim and \Bztokpi}
\label{sec:hhMethod}

We reconstruct the candidate decays
$B_{\rm rec} \to h^+ h^{\prime-}$  from pairs of oppositely-charged tracks 
that are consistent with originating 
from a common decay point with a $\chi^2$ probability of at least 0.001.
The remaining particles are examined to infer whether 
the other $B$ meson in the event (\Btag) decayed as a $\Bz$ or $\Bzb$ (flavor tag).
We perform an unbinned extended ML fit 
to separate $\Bztopippim$ and $\Bztokpi$ decays
and determine simultaneously their 
\CP-violating asymmetries \spipi, \cpipi, and 
\begin{equation}
\akpi = {\BR(B\to K^-\pi^+) - \BR(B\to K^+\pi^-) \over 
         \BR(B\to K^-\pi^+) + \BR(B\to K^+\pi^-)},
\end{equation}
as well as the signal and background yields and PDF parameters.  
The fit uses $\theta_{\rm C}$, \dedx, \de, \mes, \fish, $B_{\rm tag}$ flavor, 
and $\deltat$ information.

The value of \de is calculated assuming that both tracks are charged
pions.  The $\Bz \to \pip \pim$ signal is described by a Gaussian
distribution for \de, with a resolution of \unit[29]{\mev}.  For each
kaon in the final state, the $\de$ peak position is shifted from zero
by an amount that depends on the kaon momentum, with an average shift
of \unit[$-45$]{\mev}.  We require
\unit[$\left|\de\right|<0.150$]{\gev}.  The wide range in $\de$ allows
us to separate \Bz decays to the four final states $\pip\pim$,
$\Kp\pim$, $\pip\Km$, and $\Kp\Km$ in a single fit.
The analysis is not optimized for measuring the $\Kp\Km$ final state,
which is treated as background.
The $\mes$ resolution is \unit[2.6]{\mevcc}. We require 
\unit[$\mes > 5.20$]{\gevcc}, with events in the large range below the 
signal peak allowing the fit to effectively determine the 
background shape parameters.

\begin{figure}[!htbp]
\begin{center}
\includegraphics[width=0.85\linewidth,clip=true]{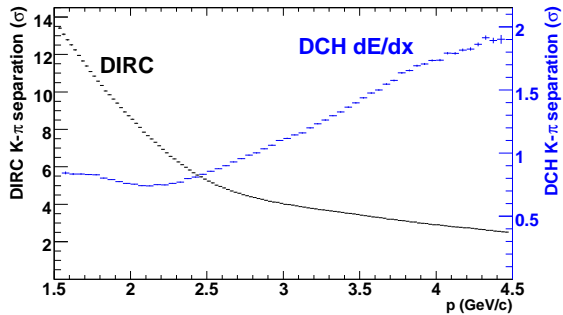}
\caption{The average expected $K-\pi$ separation, in units of uncertainty,
provided by the DIRC angle $\theta_{\rm C}$ and DCH $dE/dx$
for kaons and pions from $\Bz \to \Kp\pim$ decays in the 
laboratory-frame polar angle range $0.35 < \theta <2.40 $, 
as a function of laboratory-frame momentum. }
\label{fig:DIRC_dEdx}
\end{center}
\end{figure}

We construct $\theta_{\rm C}$ PDFs
for the pion and kaon hypotheses, and \dedx\ PDFs for the 
pion, kaon, and proton hypotheses, separately for each charge. 
The $K$--$\pi$ separations provided by $\theta_{\rm C}$ and \dedx\ are complementary:  
for $\theta_{\rm C}$, the separation varies from $2.5 \sigma$ 
at \unit[4.5]{\gevc} to $13 \sigma$ at 
\unit[1.5]{\gevc}, while for \dedx\ it varies from less than 
$1.0\sigma$ at \unit[1.5]{\gevc} to $1.9 \sigma$ at \unit[4.5]{\gevc} 
(Fig.~\ref{fig:DIRC_dEdx}). For more details, see Ref.~\cite{BaBarPRL2007}.

\begin{table}[!htbp]
\caption{ Average tagging efficiency $\epsilon$, average mistag
fraction $w$, mistag fraction difference $\Delta w = w(\Bz) -
w(\Bzb)$, and effective tagging efficiency $Q$ for signal events in
each tagging category (except the untagged category).  
}

\begin{center}
\begin{tabular}{cr@{~}c@{~}lr@{~}c@{~}lr@{~}c@{~}lr@{~}c@{~}l} \hline\hline
Category & \multicolumn{3}{c}{$\epsilon\,(\%)$} & \multicolumn{3}{c}{$w\,(\%)$} & \multicolumn{3}{c}{$\Delta w\,(\%)$} &
\multicolumn{3}{c}{$Q\,(\%)$} \rule[-2mm]{0mm}{6mm} \\\hline
{\tt Lepton}    & $8.96  $&$ \pm $&$ 0.07  $&$ 2.9  $&$ \pm $&$ 0.3 $&$  0.2 $&$ \pm $&$ 0.5 $&$ 7.95  $&$ \pm $&$ 0.11$\\
{\tt Kaon\,I}   & $10.81 $&$ \pm $&$ 0.07  $&$ 5.3  $&$ \pm $&$ 0.3 $&$  0.0 $&$ \pm $&$ 0.6 $&$ 8.64  $&$ \pm $&$ 0.14$\\
{\tt Kaon\,II}  & $17.18 $&$ \pm $&$ 0.09  $&$ 14.5 $&$ \pm $&$ 0.3 $&$  0.4 $&$ \pm $&$ 0.6 $&$ 8.64  $&$ \pm $&$ 0.17$\\
{\tt Kaon\,Pion}& $13.67 $&$ \pm $&$ 0.08  $&$ 23.3 $&$ \pm $&$ 0.4 $&$ -0.6 $&$ \pm $&$ 0.7 $&$ 3.91  $&$ \pm $&$ 0.12$\\
{\tt Pion}      & $14.19 $&$ \pm $&$ 0.08  $&$ 32.6 $&$ \pm $&$ 0.4 $&$  5.1 $&$ \pm $&$ 0.7 $&$ 1.73  $&$ \pm $&$ 0.09$\\
{\tt Other}     & $9.55  $&$ \pm $&$ 0.07  $&$ 41.5 $&$ \pm $&$ 0.5 $&$  3.8 $&$ \pm $&$ 0.8 $&$ 0.28  $&$ \pm $&$ 0.04$\\
\hline
Total           &        &       &       &        &       &       &        &       &       &$ 31.1 $&$ \pm $&$ 0.3$ \rule[-2mm]{0mm}{6mm} \\\hline\hline
\end{tabular}
\end{center}
\label{tab:tagging}
\end{table}

We use a multivariate technique~\cite{ref:sin2betaPRL04} to determine the
flavor of the $\Btag$.  Separate neural networks are
trained to identify leptons from \B\ decays, kaons from $D$ decays,
and soft pions from $D^*$ decays.  Events
are assigned to one of seven mutually exclusive tagging categories
(one category being untagged events) based on the estimated average mistag
probability and the source of the tagging information.  
The quality of tagging is expressed in
terms of the effective efficiency $Q = \sum_k \epsilon_k(1-2w_k)^2$,
where $\epsilon_k$ and $w_k$ are the efficiencies and mistag
probabilities, respectively, for events tagged in category $k$. The
difference between the mistag probabilities for $\Bz$ and $\Bzb$ mesons
is given by $\Delta w = w_{\Bz} -
w_{\Bzb}$.  Table~\ref{tab:tagging} summarizes the tagging performance
measured in a large data sample of fully-reconstructed neutral \Bflav\ decays to
$D^{(*)-}(\pip,\, \rho^+,\, a_1^+)$~\cite{rho-a1}.

The time difference $\deltat = \Delta z/\beta\gamma c$ is obtained
from the known boost of the $\epem$ system ($\beta\gamma = 0.56$) and
the measured distance $\Delta z$ along the beam ($z$) axis between the
\Brec\ and \Btag\ decay vertices.  
A description of the inclusive reconstruction of the
\Btag{} vertex is given in Ref.~\cite{ref:BaBarsin2beta}.
We require
\unit[$\left|\deltat\right|<20$]{\ps} and \unit[$\sigma_{\deltat} < 2.5$]{\ps}, where
$\sigma_{\deltat}$ is the uncertainty on $\deltat$, estimated separately for
each event. The signal \deltat PDF for $\Bztopippim$ is given by
\begin{widetext}
\begin{equation}
f^{\pm}_{k}(\deltat_{\rm meas}) = \frac{e^{-|\deltat|/\tau}}{4\tau}
           \Bigl\{ ( 1 \mp \Delta w )  
                    \pm ( 1-2w_{k} ) \bigl[  \spipi \sin{(\deltamd\deltat)} -  \cpipi\cos{(\deltamd\deltat)}    
                   \bigr]   
           \Bigr\} \otimes R(\deltat_{\rm meas} - \deltat),
\end{equation} 
\end{widetext}
where $f^{+}_{k}$ ($f^{-}_{k}$) indicates a \Bz (\Bzb) flavor tag and
the index $k$ indicates the tagging category.  The resolution function
$R(\deltat_{\rm meas} - \deltat)$ for signal candidates is a sum of three 
Gaussian functions,
identical to the one described in Ref.~\cite{ref:BaBarsin2beta}, with
parameters determined from a fit to the \Bflav\ sample, which includes
events in all seven tagging categories.  The background $\deltat$
distribution is modeled as the sum of three Gaussians, with
parameters, common for all
tagging categories, determined simultaneously with the \CP violation
parameters in the ML fit to the $B_{\rm rec} \to h^+ h^{\prime-}$ sample.

The ML fit PDF includes 28 components. 
Of these, 24 components correspond to 
\Bz\ signal decays and background events with
the final states $\pip\pim$, $\Kp\pim$, $\Km\pip$, and $\Kp\Km$,
where either the positively-charged track, the negatively-charged track,
or both have good DIRC information ($2 \times 4 \times 3 = 24$ 
components). Four additional components correspond to $\proton\pim$, $\proton\Km$, $\pip\antiproton$
and $\Kp\antiproton$ background events, where the (anti)proton
has no DIRC information. The $\Kpm\pimp$ event yields $n_{\Kpm\pimp}$ 
are parameterized 
in terms of the asymmetry $\akpi^{\rm raw}$ 
and average yield $n_{K\pi}$ as
$n_{\Kpm\pimp}=n_{K\pi}\left(1\mp \akpi^{\rm raw} \right)/2$. All other event yields are 
products of the fraction of events in each tagging category, taken from
\Bflav\ events, and the total event yield.  The background PDFs
are a threshold function~\cite{ref:argus}
for \mes and a second-order polynomial for \de.  The \fish
PDF is a sum of two asymmetric Gaussians for both  signal and background.  
We use large samples of simulated $B$ decays to investigate the effects of 
backgrounds from other $B$ decays on the determination of the \CP-violating
asymmetries in \Bztopippim and \Bztokpi, and find them to be negligible.

\subsubsection{\boldmath \Bztopizpiz}
\label{sec:pizpizMethod}

\Bztopizpiz events are identified with an ML fit to the variables
\mes, \de, and the output \emph{NN} of the event-shape neural
network.  We require \unit[$\mes>5.20$]{\gevcc} and \unit[$|\de| <
0.2$]{\gev}.  Since tails in the EMC response produce a correlation
between \mes and \de, a two-dimensional binned PDF, derived from the
signal MC sample, is used to describe signal PDF. The \emph{NN}
distribution is divided into ten bins (with each bin approximately equally
populated by signal events) and described by a nine-bin step-function PDF
with values taken from the MC and fixed in the fit. \Bflav\ data
are used to verify that the MC accurately reproduces the \emph{NN}
distribution.  The \qq background PDFs are a threshold
function~\cite{ref:argus} for \mes, a second-order polynomial for \de,
and a parametric step function for \emph{NN}.  For \qq events,
\emph{NN} is not distributed uniformly across the bins but rises
sharply toward the highest bins.  We see a small correlation 
of 2.5\%
between
the shape parameter of the \mes threshold function and the \emph{NN}
bin number, and this relation is taken into account in the fit. All
\qq background PDF-parameter values are determined by the ML fit.

The decays \Bptorhoppiz and $\Bztokspiz\ (\KS\to\piz\piz)$ add $71\pm
10$ background events to \Bztopizpiz and are included as an
additional component in the ML fit.  We model these \B-decay
background events with a two-dimensional binned PDF in \mes and \de, and
with a step function for \emph{NN}. The shapes of
these PDFs are taken from MC simulation, and their event yields
and asymmetries are fixed in the fit and are later varied to evaluate
systematic uncertainties.

The time-integrated \CP asymmetry is measured by the \B-flavor
tagging algorithm described above.  The fraction of events in each tagging
category is constrained to the corresponding fraction determined from
MC simulation.  The PDF event yields for the \Bztopizpiz signal are
given by the expression
\begin{equation}
n_{\piz\piz, k} = \frac{1}{2} f_{k} N_{\piz\piz} \Bigl[ 1 - s_j
  (1-2\chi)(1-2w_k) \cpizpiz \Bigr],
\end{equation}
where $f_k$ is the fraction of events in tagging category $k$,
$N_{\piz\piz}$ is the number of \Bztopizpiz candidate decays, 
$\chi$ 
is the time-integrated \Bz mixing probability~\cite{pdg}, $s_j=+1(-1)$
when the \Btag\ is a \Bz (\Bzb),
and 
\begin{equation}
\label{eq:directCP}
  \cpizpiz = \frac{|A^{00}|^{2} - |\Abar^{00}|^{2}}{|A^{00}|^{2} +
    |\bar{A}^{00}|^{2}}
\end{equation}
is the direct \CP asymmetry in \Bztopizpiz.

\subsection{\boldmath\Bztokspiz}
\label{sec:kspizMethod}

CP-violation parameters for \Bztokspiz\ have been reported in
Ref.~\cite{Aubert:2008ad}. Here we describe the measurement of the
branching fraction for this mode.

For each \Bztokspiz\ candidate, two independent kinematic variables
are computed.  The first variable is the invariant mass $\mb$ of the
\Brec. The second variable is the invariant (missing) mass $\mmiss$ of
the \Btag, computed from the magnitude of the difference between the
four-momentum of the initial $\epem$ system and that of the \Brec,
after applying a $\Bz$-mass constraint to the \Brec~\cite{ref:kspi0prd05}.  For
signal decays, $\mb$ and $\mmiss$ peak near the \Bz mass with 
resolutions of about 36 and \unit[5.3]{\mevcc}, respectively.
Since the linear correlation coefficient between $\mb$ and $\mmiss$
vanishes, these variables yield better separation of signal from
background than \mes and \de.
Both the $\mb$ and $\mmiss$ distributions exhibit a low-side tail due to 
leakage of energy out of the EMC.  We select candidates
within the ranges \unit[$5.13<\mb<5.43$]{\gevcc} and 
\unit[$5.11<\mmiss<5.31$]{\gevcc}, which include a signal peak and a
``sideband'' region for background characterization. In
events with more than one reconstructed
candidate (0.8\% of the total), we select the candidate with the smallest
$\chi^2=\sum_{i=\piz,\KS} (m_i-m'_i)^2/\sigma^2_{m_i}$, where $m_i$
($m'_i$) is the measured (nominal) mass and $\sigma_{m_i}$ is the
estimated uncertainty on the measured mass of particle $i$.

We exploit topological observables, computed in the c.m.\ frame, to
discriminate jet-like $\epem \to \qqbar$ events from the
nearly spherical \BB{} events.  
In order to reduce the number of background events, we require $L_2/L_0<0.55$, 
where $L_j\equiv\sum_i |{\bf p}^*_i| \cos^j \theta^*_i$ and 
$\theta^*_i$ are computed with respect to the sphericity
axis~\cite{Bjorken:1969wi} of the \Brec\ candidate.
Taking advantage of the fact that signal events follow a $1-\cos^2\thetacms$
distribution while the background is flat, we select events with
$|\costhetacms| < 0.9$. 
Using a full detector simulation, we estimate that our selection
retains $(34.2 \pm 1.2)\%$ of the signal events, where the uncertainty
includes both statistical and systematic contributions.  The selected
sample of \Bztokspiz{} candidates is dominated by random $\KS\piz$
combinations from $\epem\to\qqbar$ events.  Using large samples
of simulated \BB{} events, we find that backgrounds from other
\B-meson decays are small, of order 0.1\%. Therefore, this type
of background is not included in the fit described below,
and this is accounted for in the evaluation of systematic uncertainties 
(see Section~\ref{sec:kspizResults}).

We extract the signal yield from an extended unbinned ML fit to 
$\mb$, $\mmiss$, $L_{2}/L_{0}$, $\cos\thetacms$, 
the flavor-tag, and the decay time and its error.
The use of tagging and decay-time information in the ML fit further improves 
discrimination between signal and background.
Since in the \Bztokspiz{} decay no charged particles originate from
the decay vertex, we compute the decay point of the \Brec{} using the
$\KS$ trajectory, obtained from the reconstructed $\KS$ decay vertex
and momentum vector, and the average $\epem$ interaction
point~\cite{ref:kspi0prd08}.
We have verified that all correlations between the fit variables
are negligible, and so construct the likelihood function as a product of
one-dimensional PDFs.  Residual correlations are taken into account in
the systematic uncertainty, as explained below.

The PDFs for signal events are parameterized based on a large sample of
fully-reconstructed $B$ decays in data and from simulated events.  For
background PDFs, we take the functional form from the
background-dominated sideband regions in the data.
The likelihood function is:
\begin{widetext}
\begin{eqnarray}
\label{eq:ml}
\lefteqn{
{\cal L}(\sf,\cf,N_{\rm S},N_{\rm B},f^g_{\rm S},f^g_{\rm B},\vec{\alpha}) 
     =\frac{e^{-(N_{\rm S}+N_{\rm B})}}{N\,!}} &&   \\
&& \times \prod_{i \in g}
    \left[ N_{\rm S} f^g_{\rm S} \epsilon^{c}_{\rm S}{\cal P}_{\rm S}(\vec{x}_i,\vec{y}_i;\sf,\cf) + 
    N_{\rm B} f^g_{\rm B} \epsilon^{c}_{\rm B} {\cal P}_{\rm B}(\vec{x}_i,\vec{y}_i;\vec{\alpha}) \right]
 \prod_{i \in  b}
    \left[ N_{\rm S} f^b_{\rm S} \epsilon^{c}_{\rm S} {\cal P}'_{\rm S}(\vec{x}_i;\cf) + 
     N_{\rm B} f^b_{\rm B} \epsilon_{\rm B}^{c} {\cal P}'_{\rm B}(\vec{x}_i;\vec{\alpha}) \right], 
\nonumber
\end{eqnarray}
\end{widetext}
where the $N$ selected events are partitioned into two subsets: the
index $i \in g$ indicates events that have \deltat information, while
$i \in b$ events do not have \deltat information. Here, $f^g_{\rm S}$ ($f^g_{\rm B}$) is the
fraction of signal (background) events that are in the subset $g$,
and $f^b_{\rm S} = 1 - f^g_{\rm S}$ ($f^b_{\rm B} = 1 - f^g_{\rm B}$)
are the corresponding signal (background) fractions in the subset $b$.
The parameter $N_{\rm S}$ ($N_{\rm B}$) is the number of signal (background) events.
The probabilities ${\cal P}_{\rm S}$ and
${\cal P}_{\rm B}$ are products of PDFs for the signal and
background hypotheses evaluated for the measurements
$\vec{x}_i=\{\mb,\;\mmiss,\;L_{2}/L_{0},\;\costhetacms,$ flavor
tag, tagging category$\}$ and
$\vec{y}_i=\{\deltat,\sigma_{\deltat}\}$.  
The corresponding PDFs for events without
$\deltat$ information are 
${\cal P}'_{\rm S}$ and
${\cal P}'_{\rm B}$. 
Detailed descriptions of ${\cal P}_{\rm S}$, 
${\cal P}_{\rm B}$, ${\cal P}'_{\rm S}$, and
${\cal P}'_{\rm B}$ are given in Ref.~\cite{Aubert:2008ad}.
The vector $\vec{\alpha}$ represents the
set of parameters that define the shapes of the PDFs.  Along with the
\CP asymmetries \sf\ and \cf, the fit extracts the yields $N_{\rm S}$ and
$N_{\rm B}$, the fraction of events $f^g_{\rm S}$ and $f^g_{\rm B}$, and the 
parameters of the background PDFs.  

\section{RESULTS AND SYSTEMATIC UNCERTAINTIES}
\label{sec:Physics}

\subsection{\boldmath \Bztopippim and \Bztokpi Results}
\label{sec:hhResults}

\begin{table}[!htbp]
\caption{  
  Results for the \Bztohh decay modes.  
  Uncertainties on the signal yields $N_{\rm sig}$ are statistical. For the \CP-violation parameters,
  the first uncertainties are statistical, and the second are  systematic.}
\label{tab:resultsB}
\begin{center}
\begin{tabular}{l|c|c}
\hline\hline
Mode        & $N_{\rm sig}$      & \CP-violation parameters       \\ 
\hline 
\Bztopippim &  $1394\pm 54$ & $\spipi = -0.68 \pm 0.10 \pm 0.03$ \\
            &               & $\cpipi = -0.25 \pm 0.08 \pm 0.02$ \\
\hline
\Bztokpi    &  $5410\pm 90$ & $\akpi  = -0.107 \pm 0.016^{+0.006}_{-0.004}$ \\
\hline
$\Bz\to K^+K^-$  &  $7 \pm 17$  & \\
\hline\hline
\end{tabular}
\end{center}
\end{table}

\begin{figure*}[!htbp]
\begin{center}
\begin{align*}
  \includegraphics[width=0.3\textwidth]{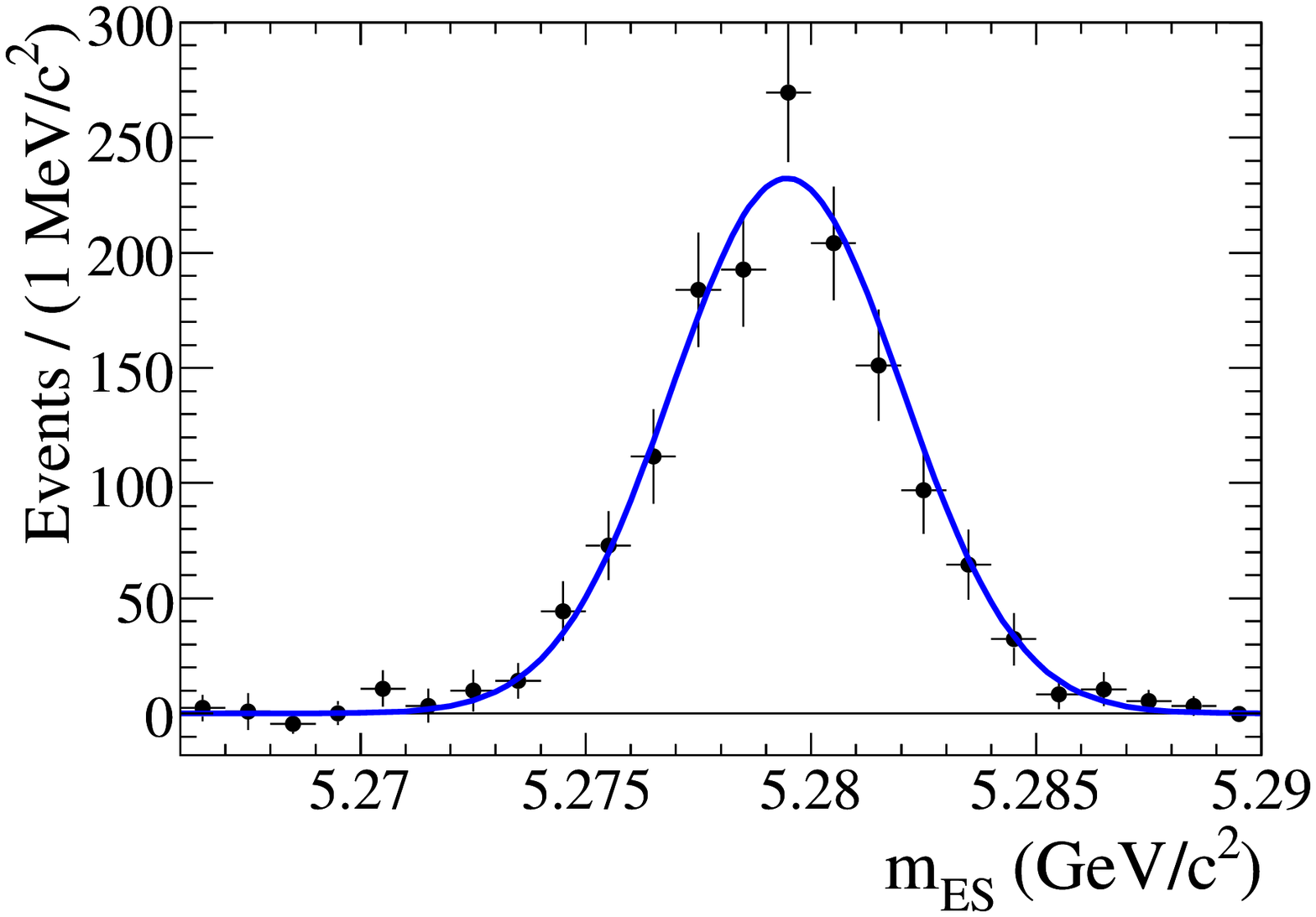} ~~~&
  \includegraphics[width=0.3\textwidth]{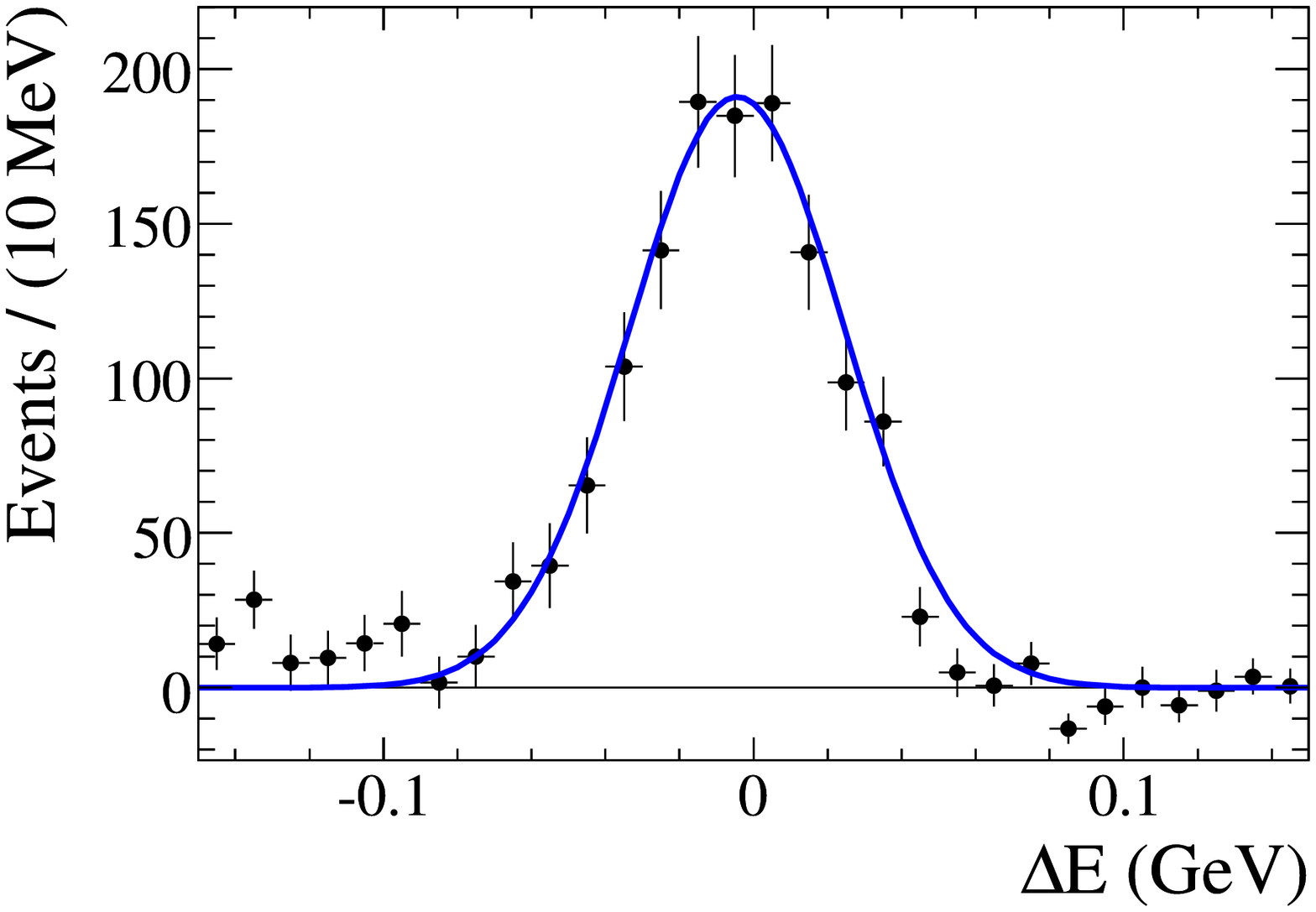} ~~~&
  \includegraphics[width=0.3\textwidth]{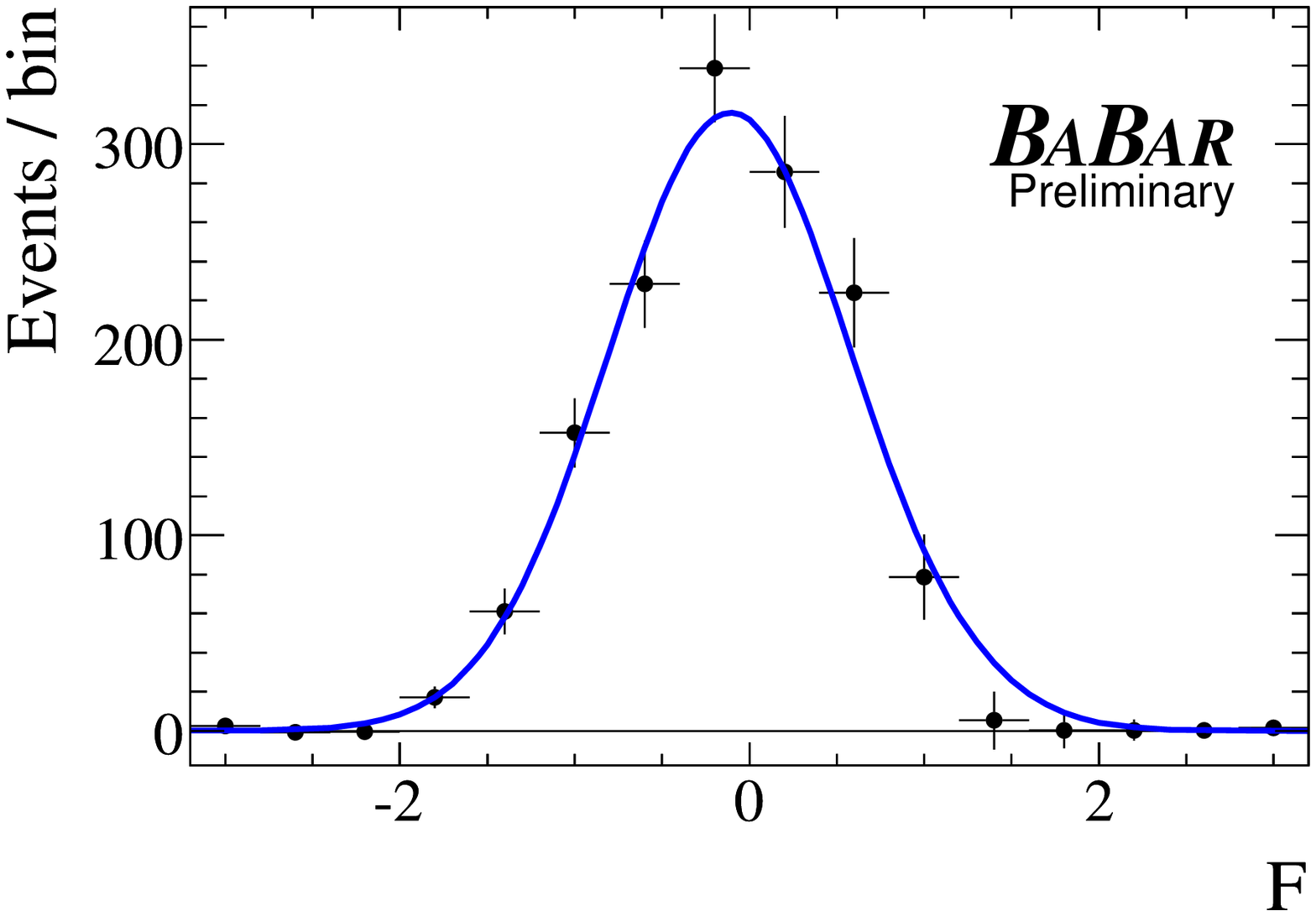} \\
\vspace*{2mm}
  \includegraphics[width=0.3\textwidth]{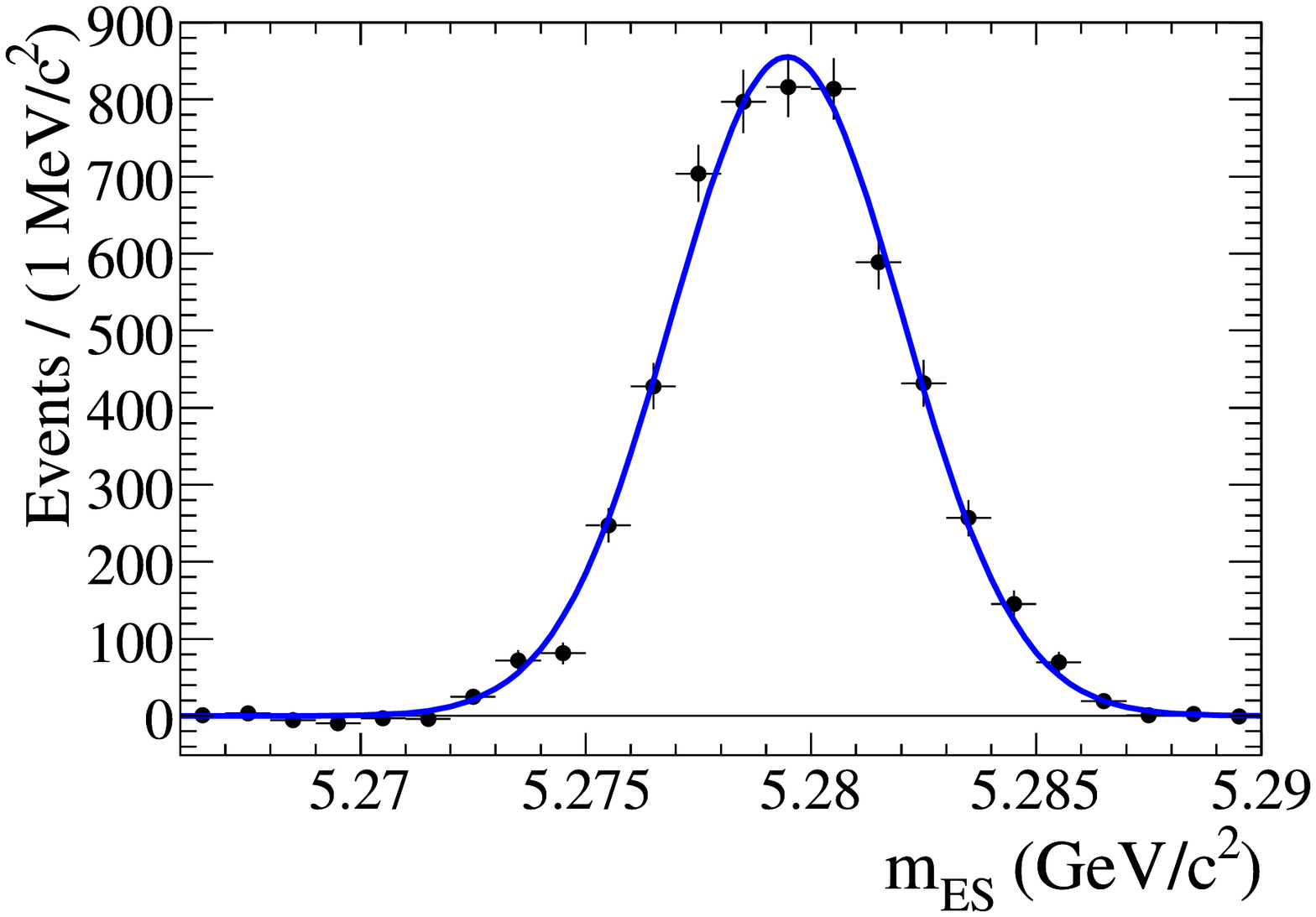} ~~~&
  \includegraphics[width=0.3\textwidth]{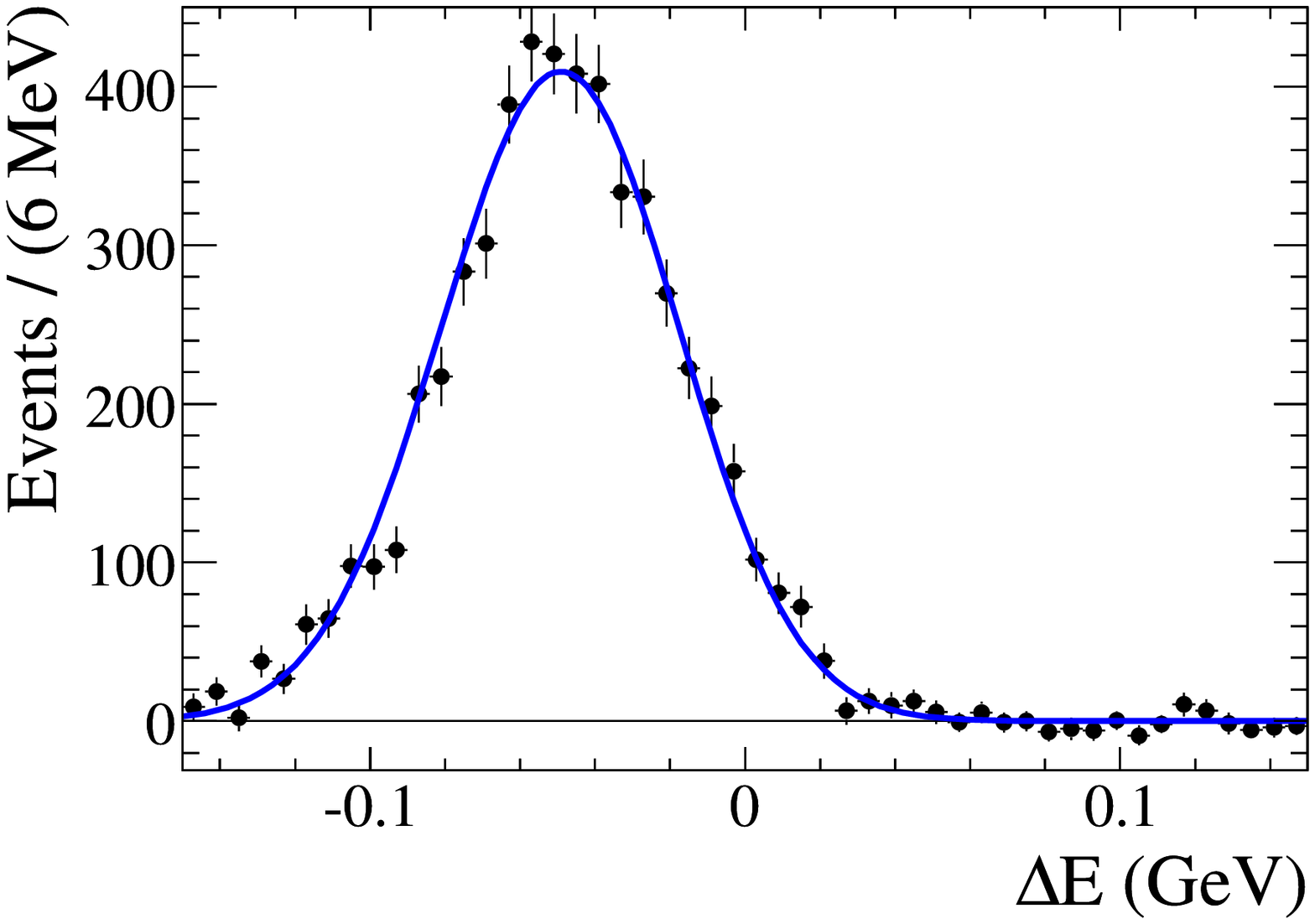} ~~~&
  \includegraphics[width=0.3\textwidth]{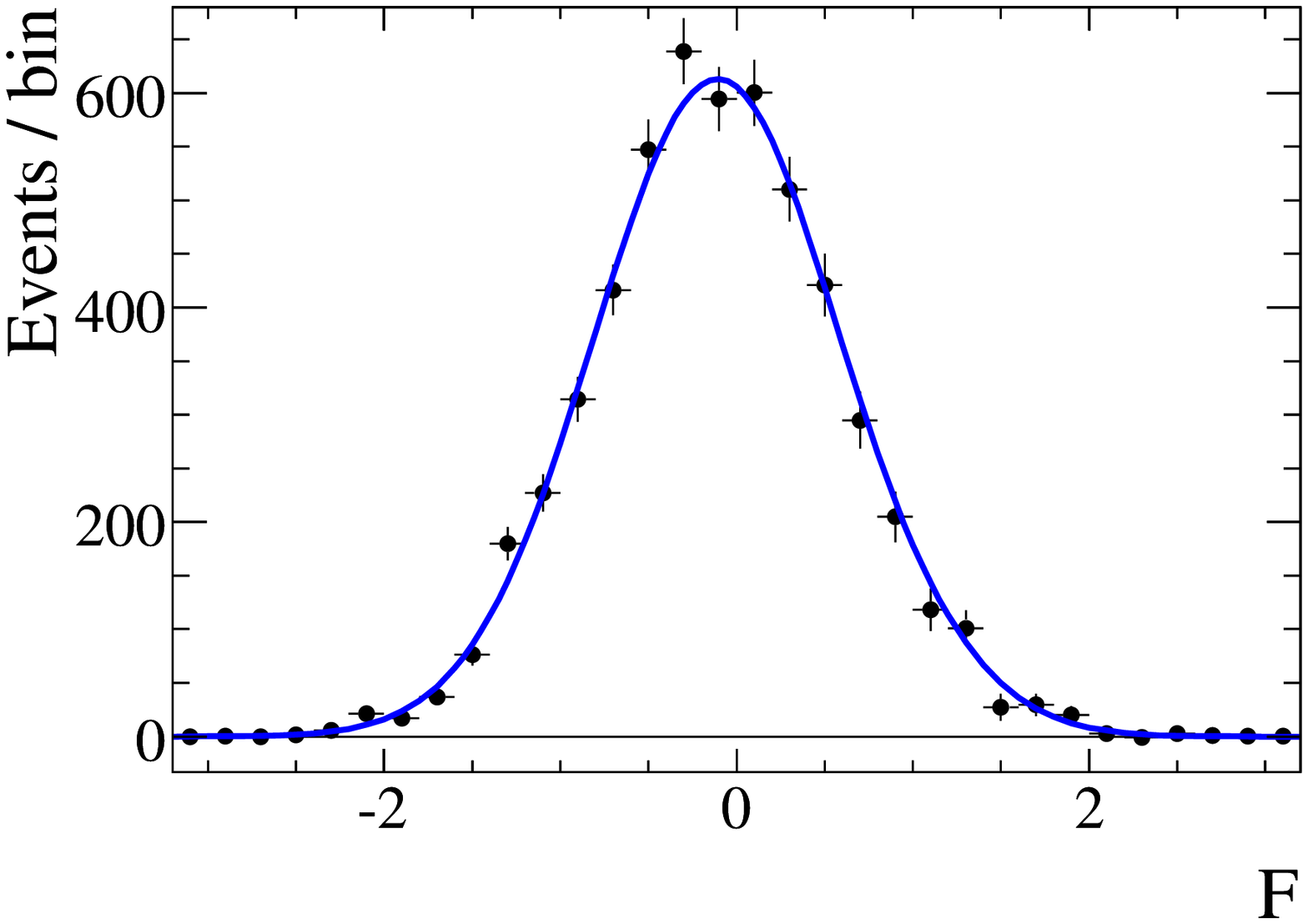} \\
\vspace*{2mm}
  \includegraphics[width=0.3\textwidth]{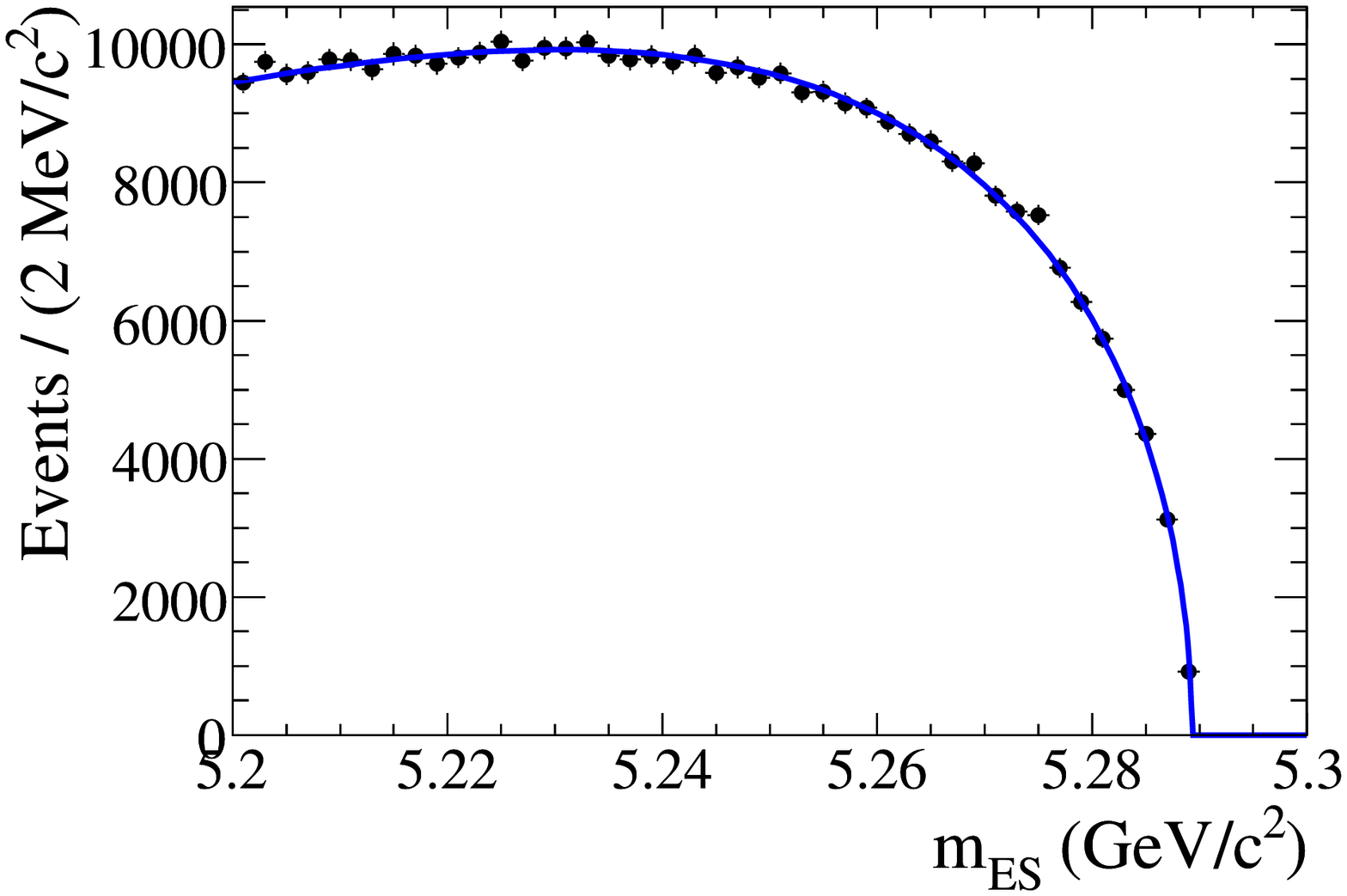} ~~~&
  \includegraphics[width=0.3\textwidth]{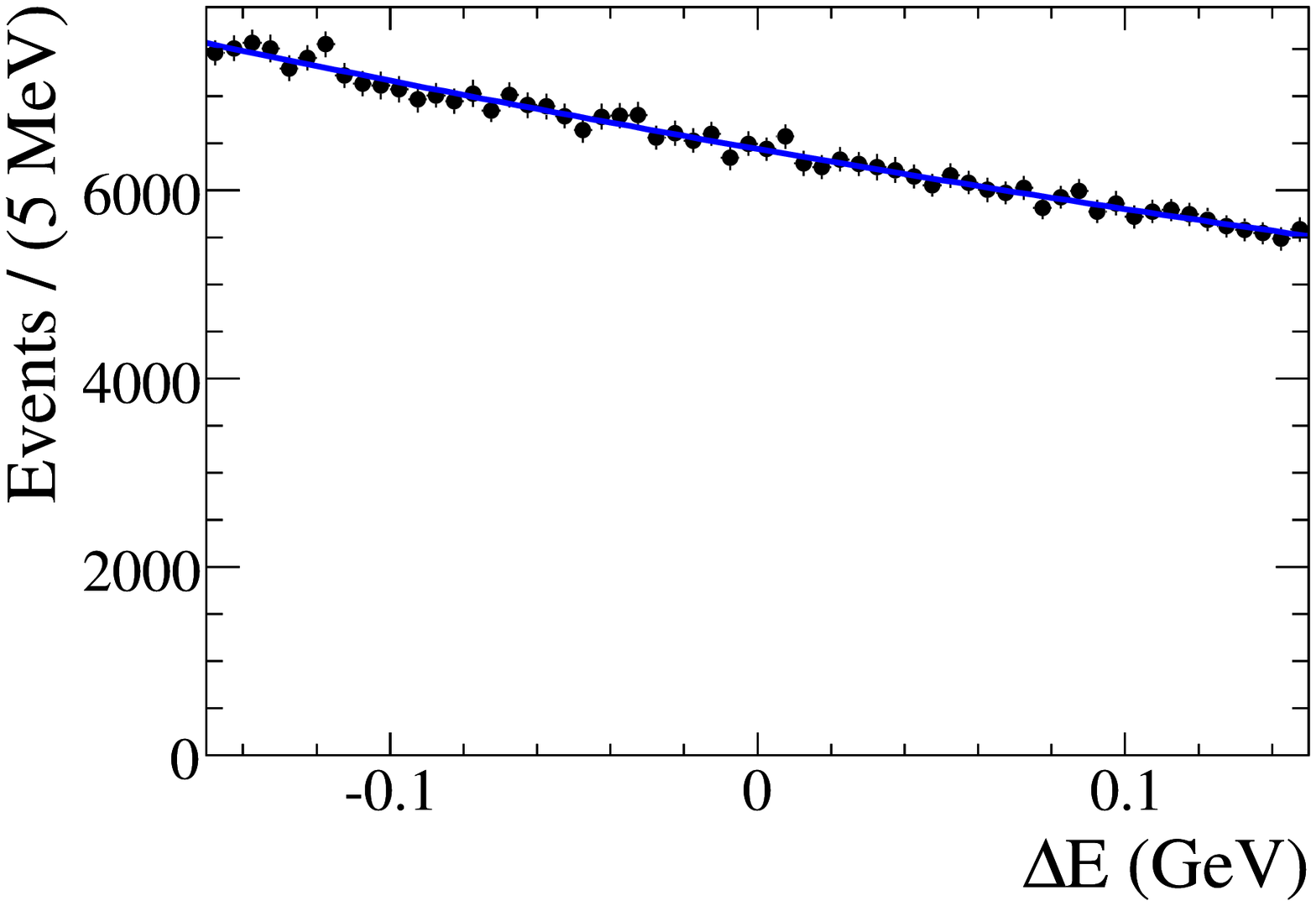} ~~~&
  \includegraphics[width=0.3\textwidth]{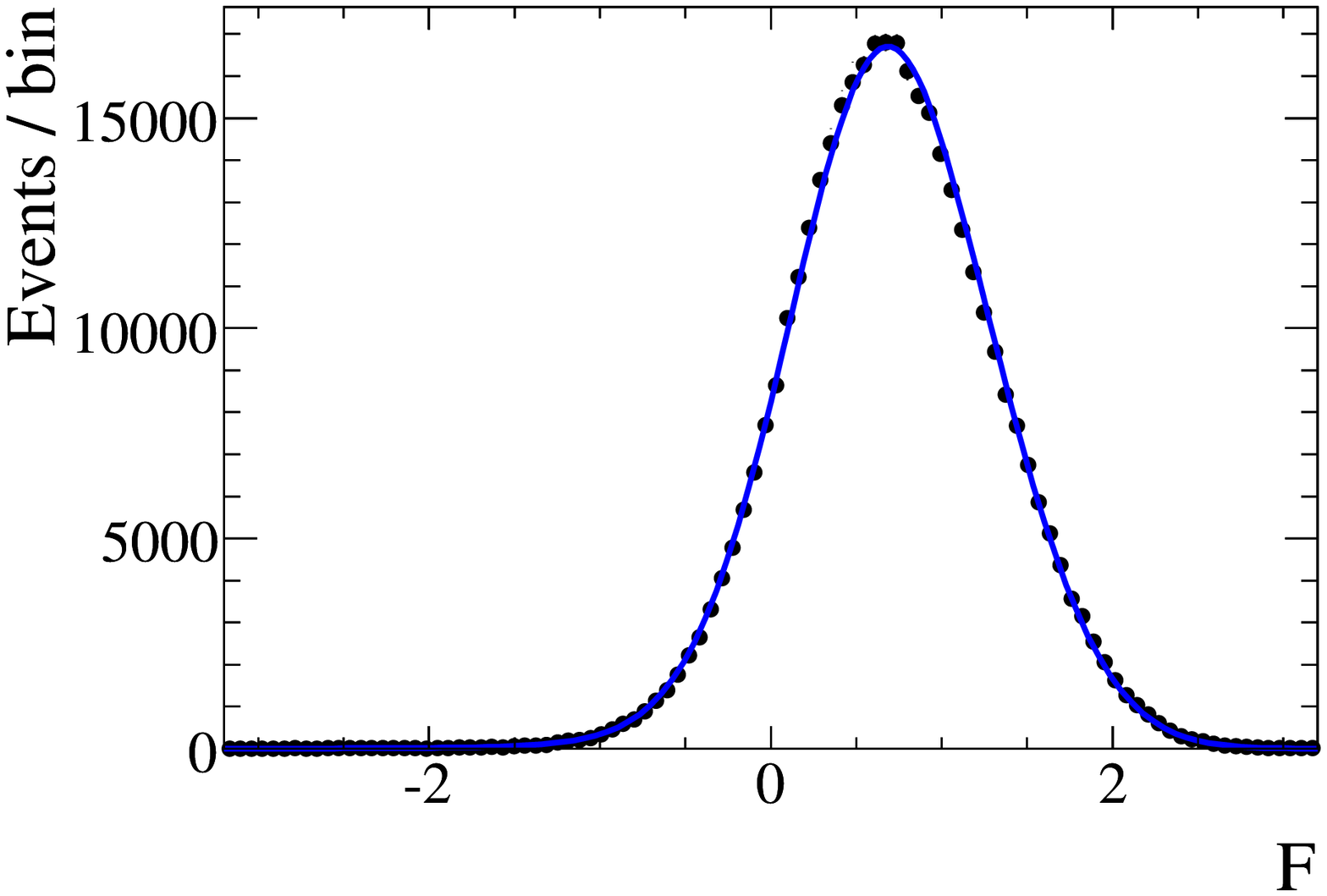} 
\end{align*}
\caption{ \sPlots of the (left column) \mes, (center column) \de, 
and (right column) Fisher discriminant \fish distributions for 
(top row) $\Bz\to\pip\pim$, (middle row) $\Bz\to\Kp\pim$,
and (bottom row) $\qq$ background candidates.
The points with error bars show 
the data, and the 
lines represent the PDFs used in the fit and reflect the fit result.
The structure to the left of the signal \DeltaE peak for $\Bz\to\pip\pim$
is consistent with the expected background from other charmless modes, which
is negligible for \unit[$\de>-0.10$]{\gev}.
In the calculation of $\de$ for $\Bz\to\Kp\pim$,
the kaon candidate is assigned the pion mass.
}
\label{fig:hhVar}
\end{center}
\end{figure*}

\begin{figure}[!htbp]
\begin{center}
  \includegraphics[width=0.7\linewidth]{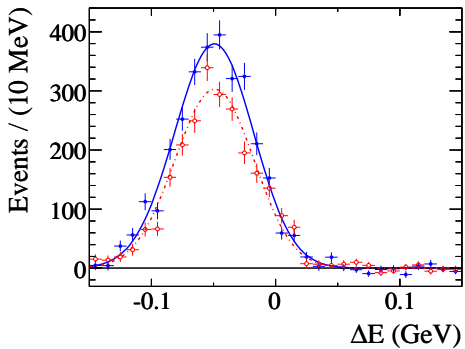}
\caption{ \sPlots of the $\de$ distribution 
for signal $\Kpm\pimp$ events, comparing (blue solid lines, filled circles) \Bz 
and (red dashed lines, empty circles) \Bzb decays. The points with error bars show 
the data, and the lines represent the PDFs used in the fits
and reflect the results of the fits.}
\label{fig:deakpi}
\end{center}
\end{figure}

\begin{figure}[!htbp]
\begin{center}
  \includegraphics[width=0.7\linewidth]{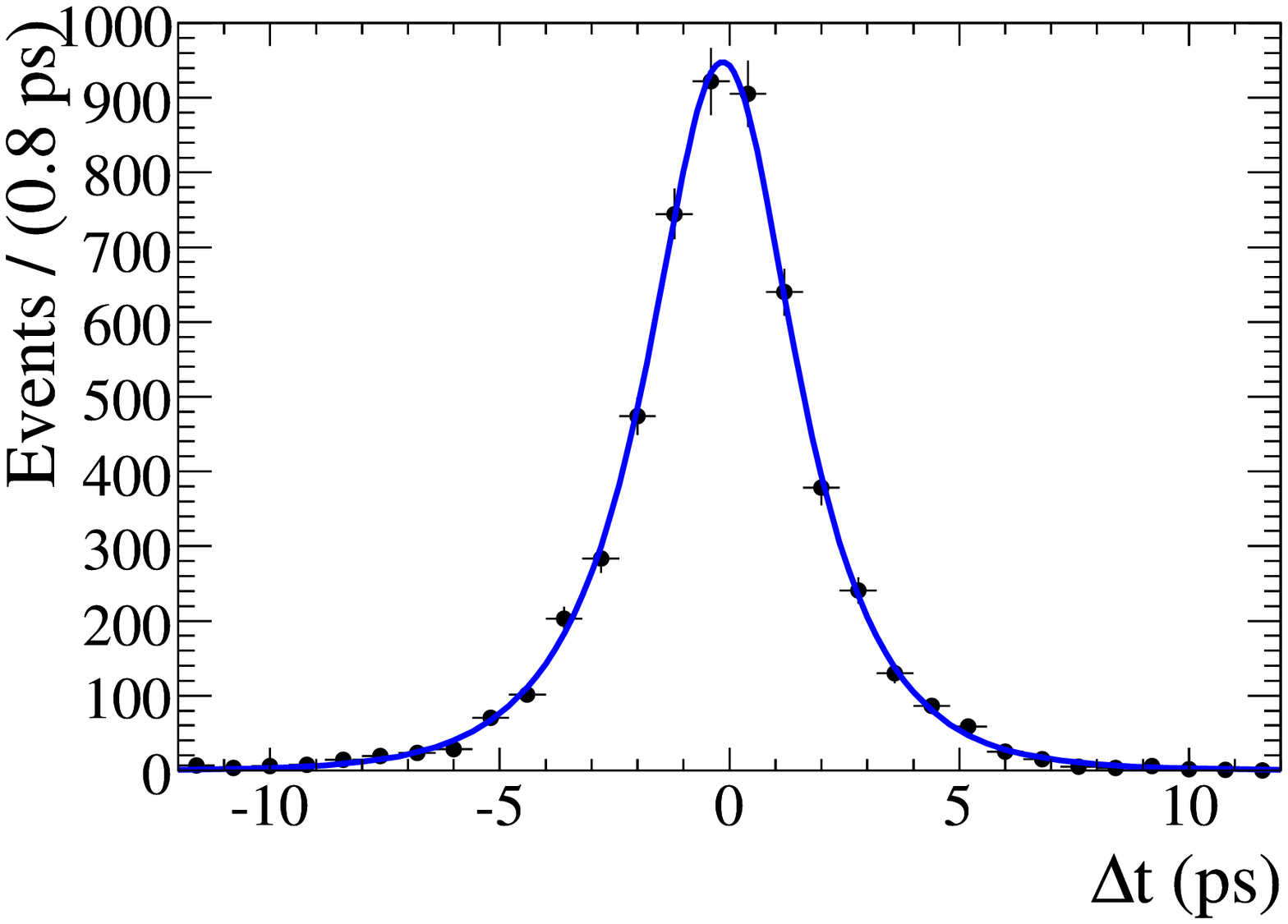}\\
\vspace*{1.5mm}
  \includegraphics[width=0.7\linewidth]{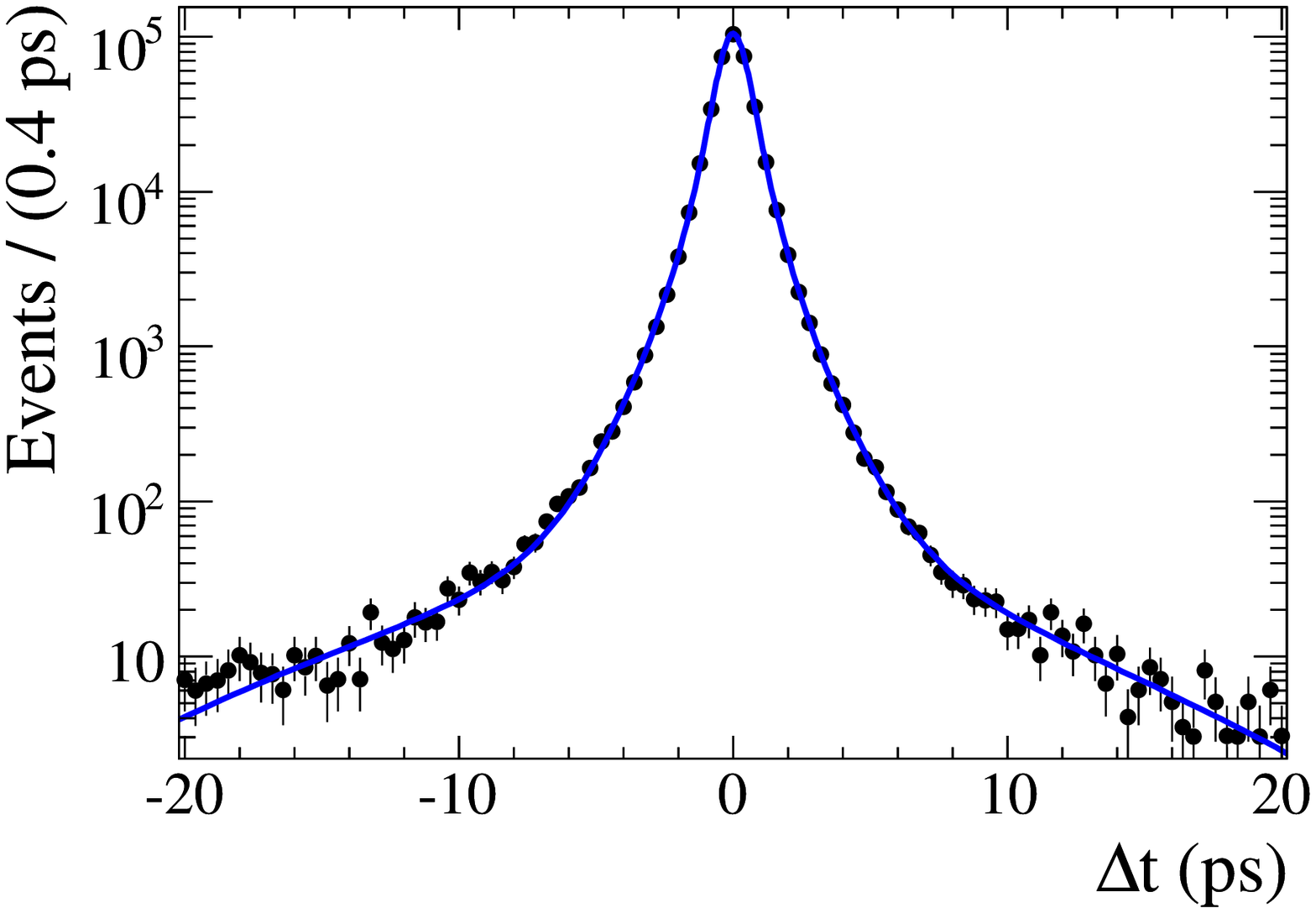}
\caption{ \sPlots of the $\deltat$ distribution for  (top)  
signal $\Kpm\pimp$ and 
(bottom) background events. The points with error bars show
the data, and the lines represent the PDFs used in the fit and 
reflect the fit result.}
\label{fig:hhdt}
\end{center}
\end{figure}

\begin{figure}[!htbp]
\begin{center}
  \includegraphics[width=0.70\linewidth]{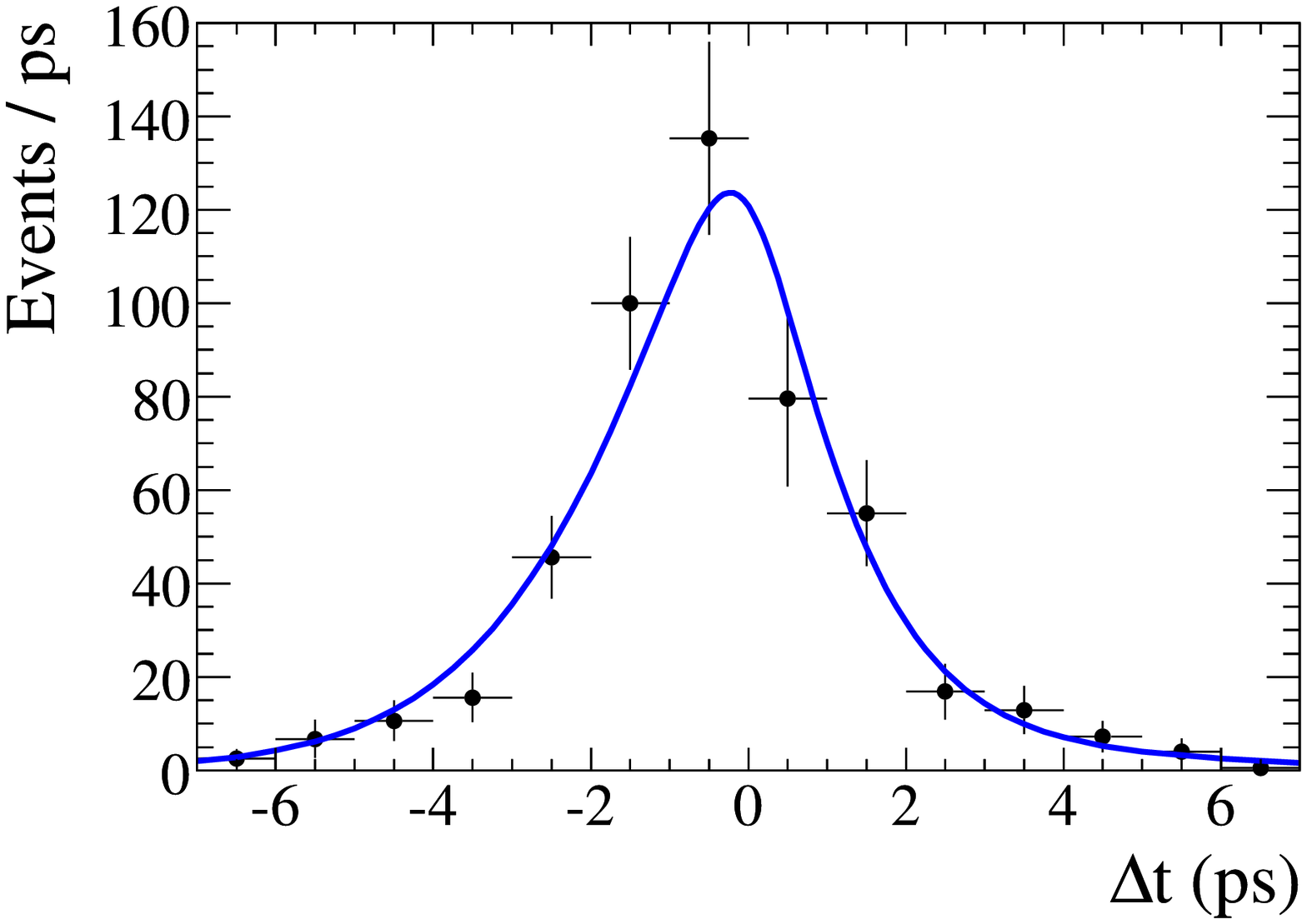} \\
\vspace*{1.5mm}
  \includegraphics[width=0.70\linewidth]{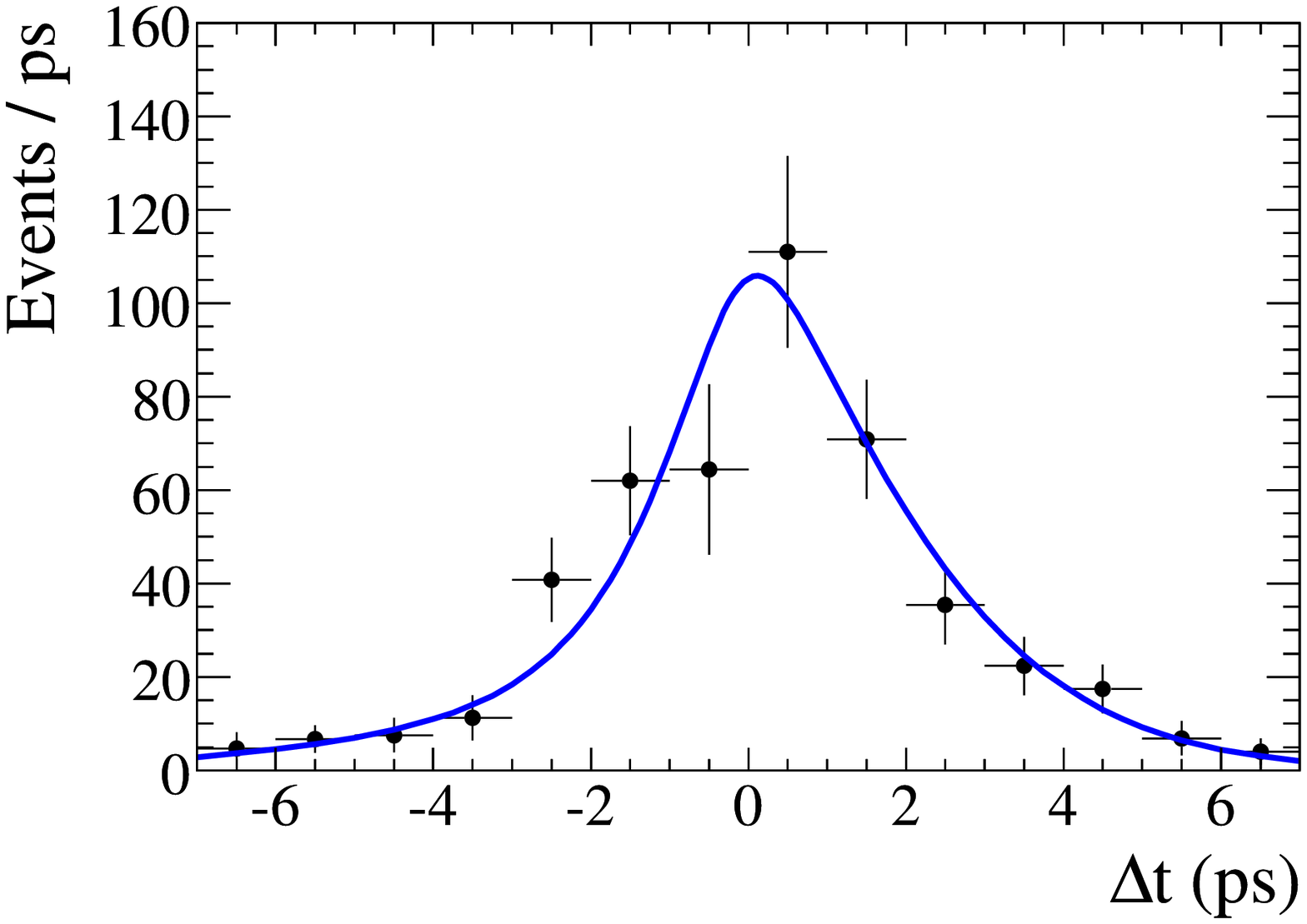} \\
\vspace*{1.5mm}\hspace*{1.0mm}
  \includegraphics[width=0.73\linewidth]{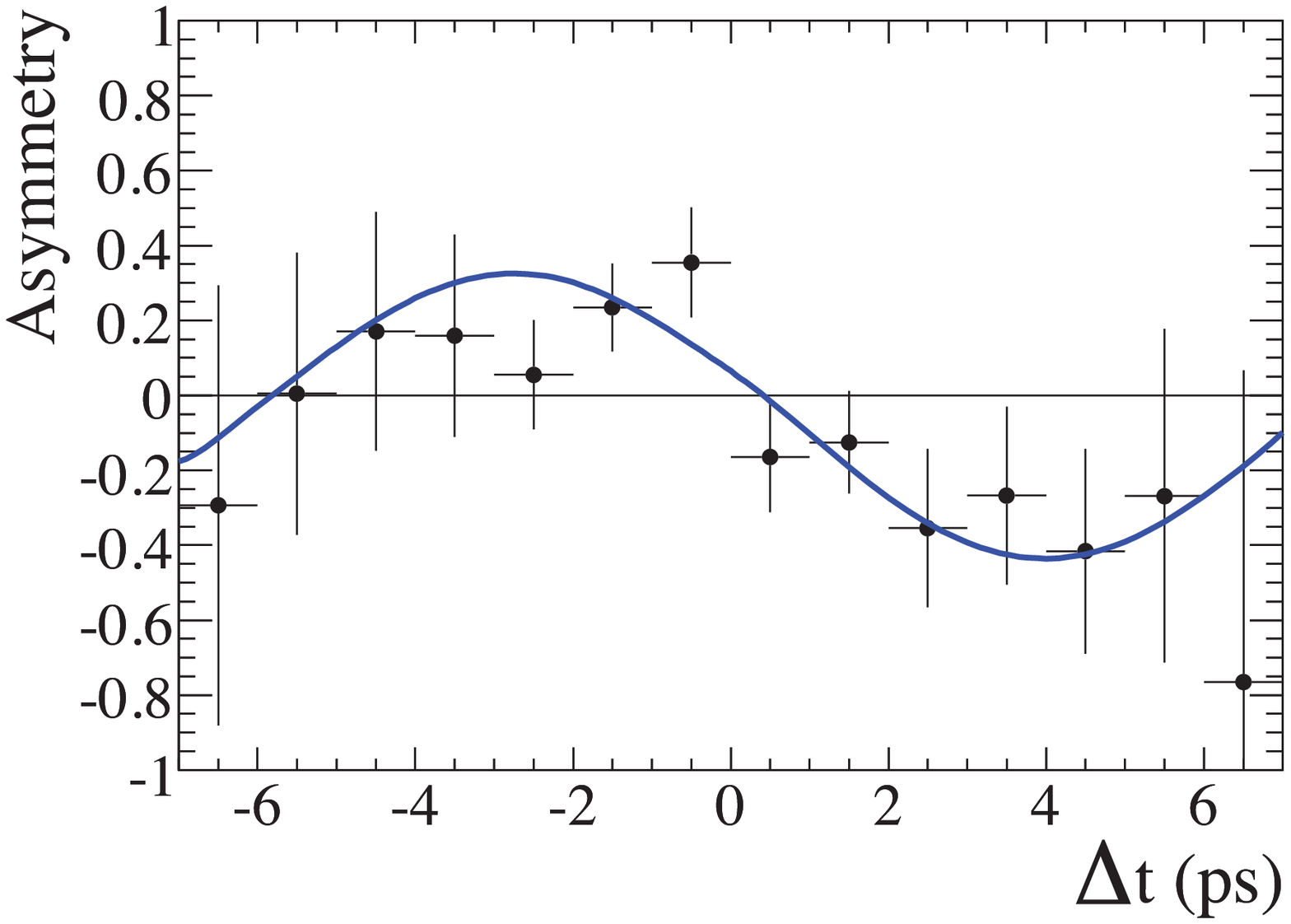}
\caption{ 
\sPlots of the 
$\deltat$ distributions for signal $\pip\pim$ events tagged as (top) $\Bz$  
or (middle) $\Bzb$, and (bottom) their asymmetry
$a(\Delta t)$, from Eq.~(\ref{eq:asymmetry}).
The points with error bars show
the data, and the lines represent the PDFs used in the fit and 
reflect the fit result.}
\label{fig:asym}
\end{center}
\end{figure}

\begin{figure}[!htbp]
\begin{center}
\includegraphics[width=0.8\linewidth]{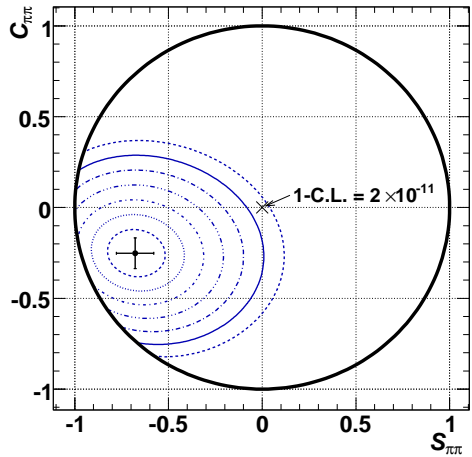}
\end{center}
\vspace{0.2cm}
\caption{ 
\spipi\ and \cpipi in $\Bz\to\pip\pim$ decays, showing the central values (point with error bars) and statistical confidence-level (C.L.) 
contours for
$1-\mathrm{C.L.} = 0.317$ $(1\sigma)$, $4.55 \times 10^{-2}$ $(2\sigma)$, $2.70 \times 10^{-3}$ $(3\sigma)$, 
$6.33 \times 10^{-5}$ $(4\sigma)$, $5.73 \times 10^{-7}$ $(5\sigma)$, $1.97 \times 10^{-9}$ $(6\sigma)$
and $2.56 \times 10^{-12}$ $(7\sigma)$,
calculated from the square root of the change in the value of
$-2\ln{\cal L}$ with respect to its value at the minimum.
The unit circle represents the physical region $\spipi^2 + \cpipi^2 \le 1$.
}
\label{fig:SCcontour}
\end{figure}

The event yields and CP-violation parameters are
listed in
Table~\ref{tab:resultsB}. The correlation coefficient between \spipi
and \cpipi is found to be $-0.056$, and 
the correlation between \cpipi and \akpi is 0.019.  
We show the $\mes$,
$\de$, and $\fish$ distribution for the 
\Btopipi, \Btokpi, and $\qqbar$ background in 
Fig.~\ref{fig:hhVar}, where the  \sPlots~\cite{ref:splots} 
weighting and background-subtraction 
technique is used to display a distribution for a particular
type of event.
%
The direct \CP asymmetry in \Bztokpi is apparent in the \de 
distributions, plotted separately for \Bz
and \Bzb decays in Fig.~\ref{fig:deakpi}.  We show the
distributions of $\deltat$ for $\Bz\to\Kpm\pimp$ signal and background decays in
Fig.~\ref{fig:hhdt}. 
In Fig.~\ref{fig:asym}, we show the distribution
of \deltat separately for \Bztopipi events tagged as $\Bz$ or $\Bzb$,
as well as the asymmetry $a(\deltat)$ of Eq.~(\ref{eq:asymmetry}).  The results for 
\spipi and \cpipi are shown in Fig.~\ref{fig:SCcontour}, 
along with confidence-level contours corresponding to 
statistical significances ranging from  $1\sigma$ to $7\sigma$.
Our measurement excludes the absence of  
\CP violation in \Bztopipi ($\spipi=0,\ \cpipi=0$) at a confidence level 
corresponding to $6.7\sigma$, including systematic uncertainties.

\begin{table}[!htbp]
\caption
{ Summary of systematic uncertainties on $\akpi$.  To address the \akpi\ bias due 
to hadronic interactions of charged kaons with the detector material, we shift the \akpi
value obtained in the fit by $+0.005$.}
\begin{center}
\begin{tabular}{lc}
\hline\hline
Source                           & $\akpi$\\
\hline
Material interactions            & $+0.005$  $-0.003$ \\
$\theta_{\rm C}$ and \dedx\ PDFs       & $0.002$  \\
Alternative DIRC parameterization & $0.002$  \\
Potential bias                & $0.001$  \\ 
\hline
Total                    & $+0.006$ $-0.004$  \\
\hline\hline
\end{tabular}
\label{tab:syst_akpi}
\end{center}
\end{table}

\begin{table}[!htbp]
\caption{ Summary of systematic uncertainties  on $\spipi$ and $\cpipi$.}
\begin{center}
\begin{tabular}{lcc}
\hline\hline
Source                     & $\spipi$ & $\cpipi$ \\
\hline
DIRC $\thetac$             & $ 0.0064   $ & $ 0.0050   $ \\
DCH $\dedx$                & $ 0.0032   $ & $ 0.0037   $ \\
Signal $\deltat$           & $ 0.0199   $ & $ 0.0055   $ \\
SVT local alignment        & $ 0.0004   $ & $ 0.0002   $ \\
Boost/detector $z$ size    & $ 0.0021   $ & $ 0.0013   $ \\
PEP-II beam spot           & $ 0.0028   $ & $ 0.0014   $ \\
\B flavor tagging          & $ 0.0146   $ & $ 0.0138   $ \\
$\deltamd$, $\tau_{\Bz}$~\cite{pdg}   & $ 0.0004   $ & $ 0.0017   $ \\
Potential bias             & $ 0.0041   $ & $ 0.0043   $ \\
\CP\ violation in $B_{\rm tag}$ decays  & $ 0.007    $ & $ 0.016    $ \\
\hline
Total                      & $ 0.027    $ & $ 0.023   $ \\
\hline\hline
\end{tabular}
\label{tab:syst_scpipi}
\end{center}
\end{table}

Systematic uncertainties for the direct \CP asymmetry $\akpi$
are listed in Table~\ref{tab:syst_akpi}. 
Here, \akpi\ is the fitted value of the $\Kmp\pipm$ event-yield asymmetry $\akpi^{\rm raw}$ 
shifted by $+0.005^{+0.005}_{-0.003}$ to account for a bias that arises 
from the difference between the cross sections of \Kp and \Km hadronic interactions
within the \babar\ detector.  We determine this bias from the MC.
The bias is independently verified with
a calculation based on the known material composition of the 
\babar\ detector~\cite{babar} and the cross sections and material properties 
tabulated in Ref.~\cite{pdg}. The corrected $\Kmp\pipm$ event-yield asymmetry in the 
background, where no observable \CP\ violation is expected, is 
$-0.005 \pm 0.004\, (\rm stat) ^{+0.005}_{-0.003}\, (\rm syst)$, consistent with zero.
Uncertainties on the $\thetac$ and $\dedx$ distributions 
are obtained from the $D^0\to K^-\pi^+$ control sample,
and contribute $0.002$ to the systematic uncertainty on $\akpi$.
An additional uncertainty of the same magnitude 
is obtained by adding a bifgurcated-Gaussian
component to the two-Gaussian $\thetac$ PDF.
We use a combination of MC events and 
parameterized experiments to test for a potential bias in the fit,
for which we estimate an uncertainty of 0.001.

Systematic uncertainties for the \CP asymmetries \spipi and \cpipi are
listed in Table~\ref{tab:syst_scpipi}. The largest uncertainties on
\spipi are due to the $\deltat$ and $B$-flavor-tagging parameters, and
are determined by varying the $\deltat$ resolution function parameters
and the flavor-tagging parameters by their uncertainties. 
The largest $\cpipi$ uncertainty is due to the effect of \CP\ violation in
the $B_{\rm tag}$ decays~\cite{Owen}.
The effect of SVT misalignment is determined by reconstructing events
with shifted alignment parameters, and the uncertainties due to the
machine boost and detector size are obtained by scaling $\deltat$ by
$1.0046$. We evaluate uncertainties due to the measurement of the beam
spot by shifting its position in the vertical direction by $20~\mu m$,
and those due to the knowledge of the $\Bz-\Bzb$ mixing frequency and
the $\Bz$ lifetime are determined by varying these parameters within
their uncertainties~\cite{pdg}.
The uncertainties due to particle identification and potential 
fit bias are evaluated as described above for $\akpi$.

\subsection{\boldmath \Bztopizpiz Results}
\label{sec:pizpizResults}

\begin{figure}[!htbp]
\begin{center}
\includegraphics[width=0.7\linewidth]{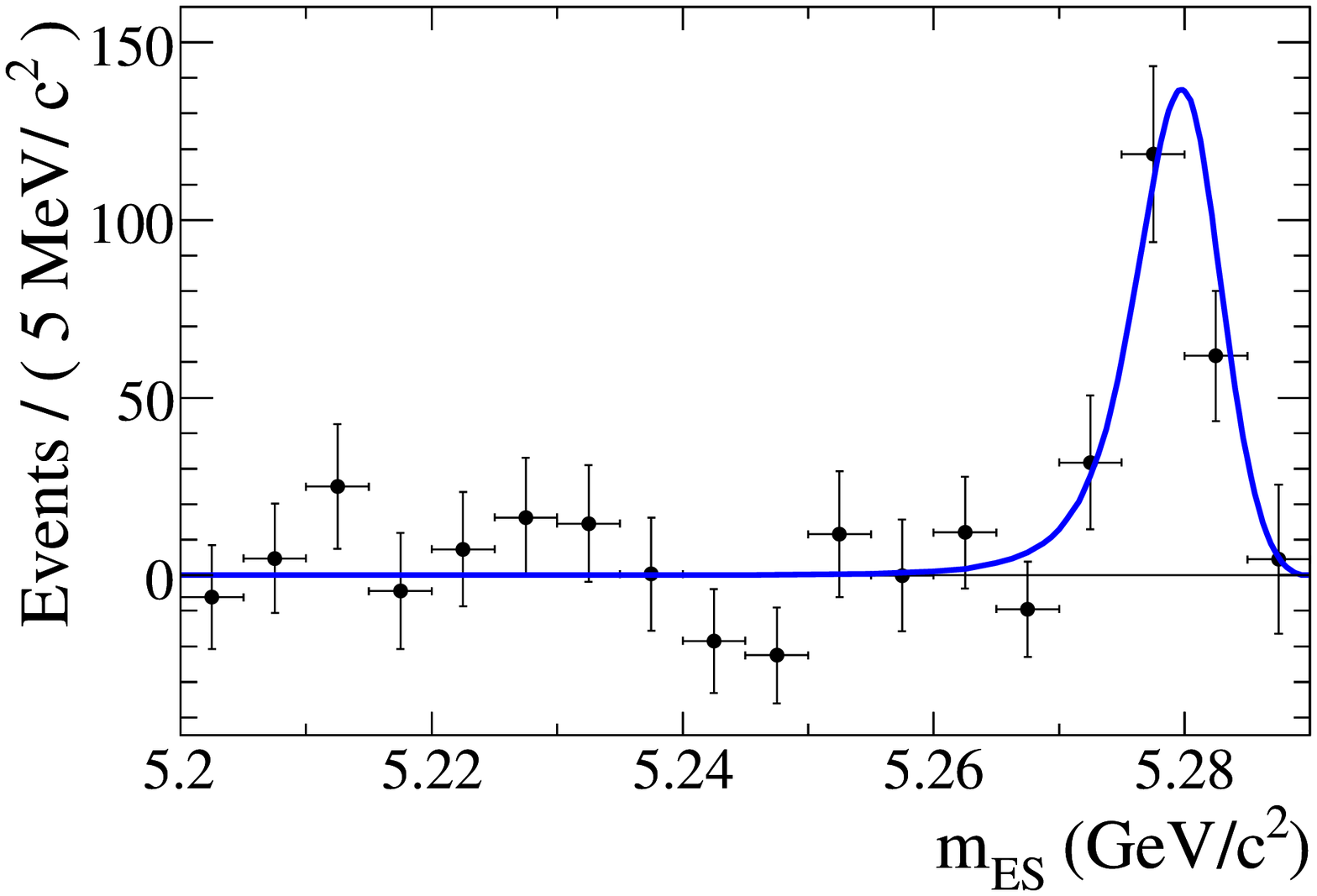}
\includegraphics[width=0.7\linewidth]{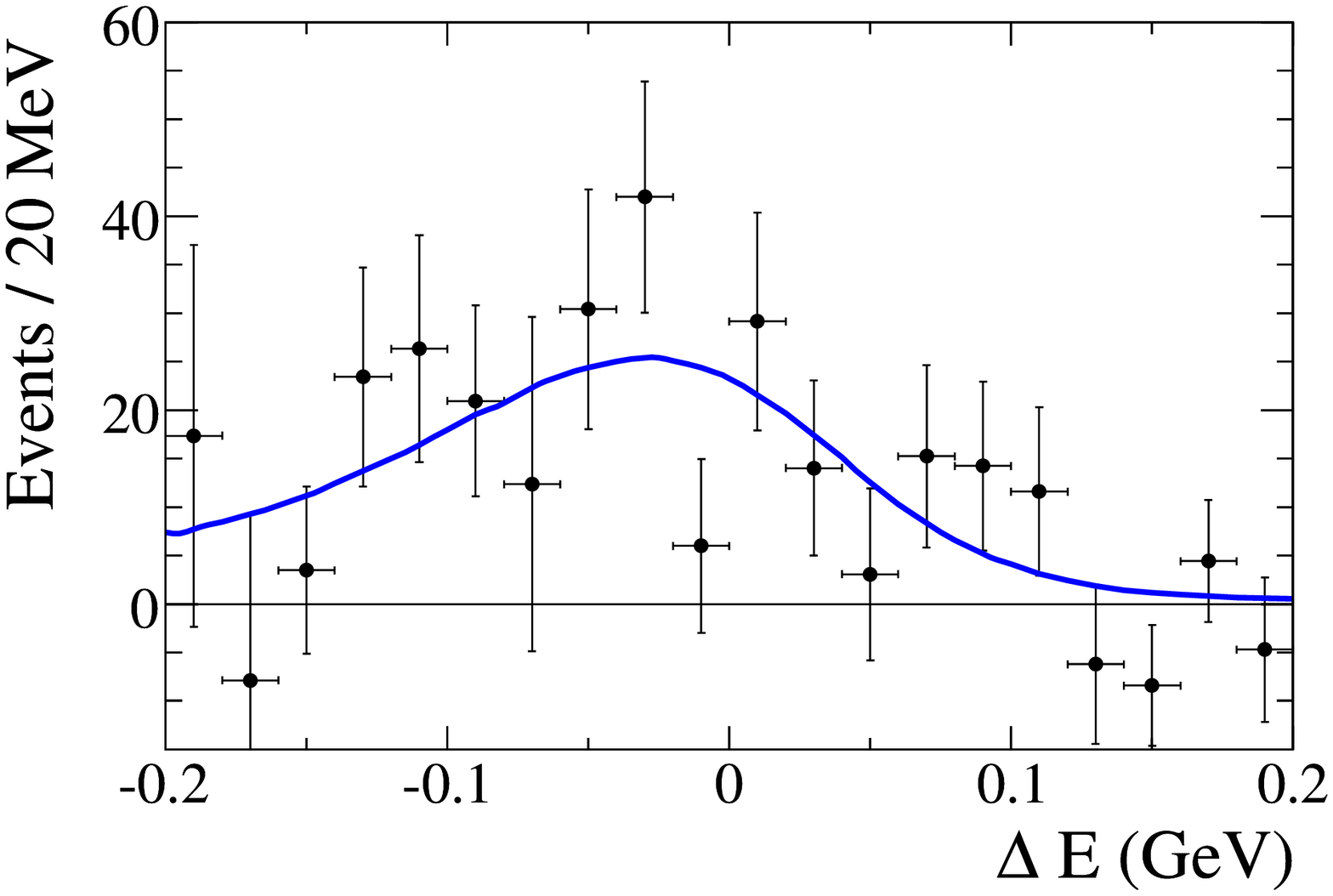}
\includegraphics[width=0.7\linewidth]{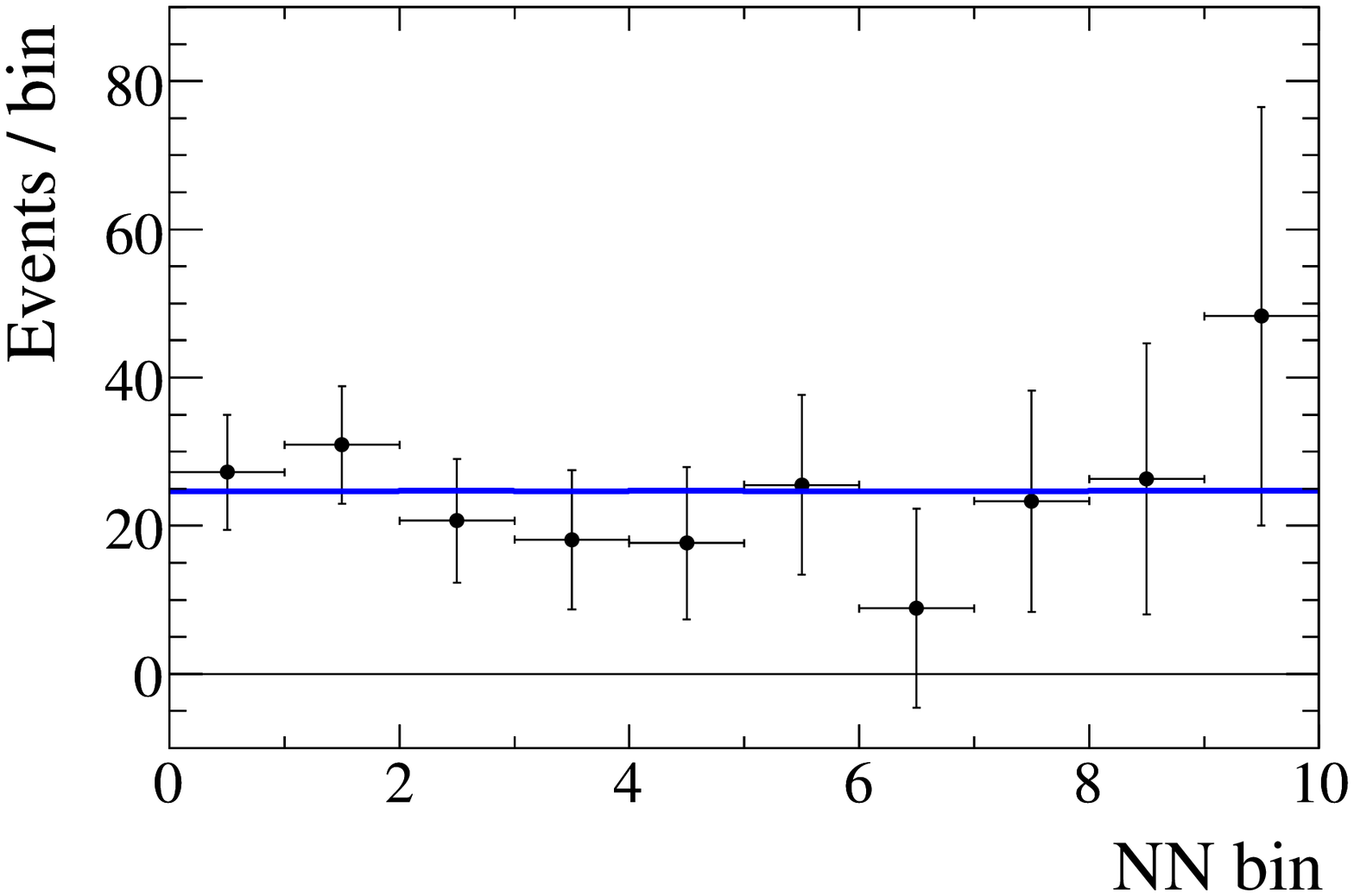}
\caption{ \Bztopizpiz signal plots with background  
subtracted using the \sPlots technique.
From top to bottom: \mes, \de, and \emph{NN}. The points with error bars show 
the data, and the line
in each plot shows the corresponding PDF.
}
\label{fig:pizpiz}
\end{center}
\end{figure}

\begin{figure}[!htbp]
\begin{center}
\includegraphics[width=0.7\linewidth]{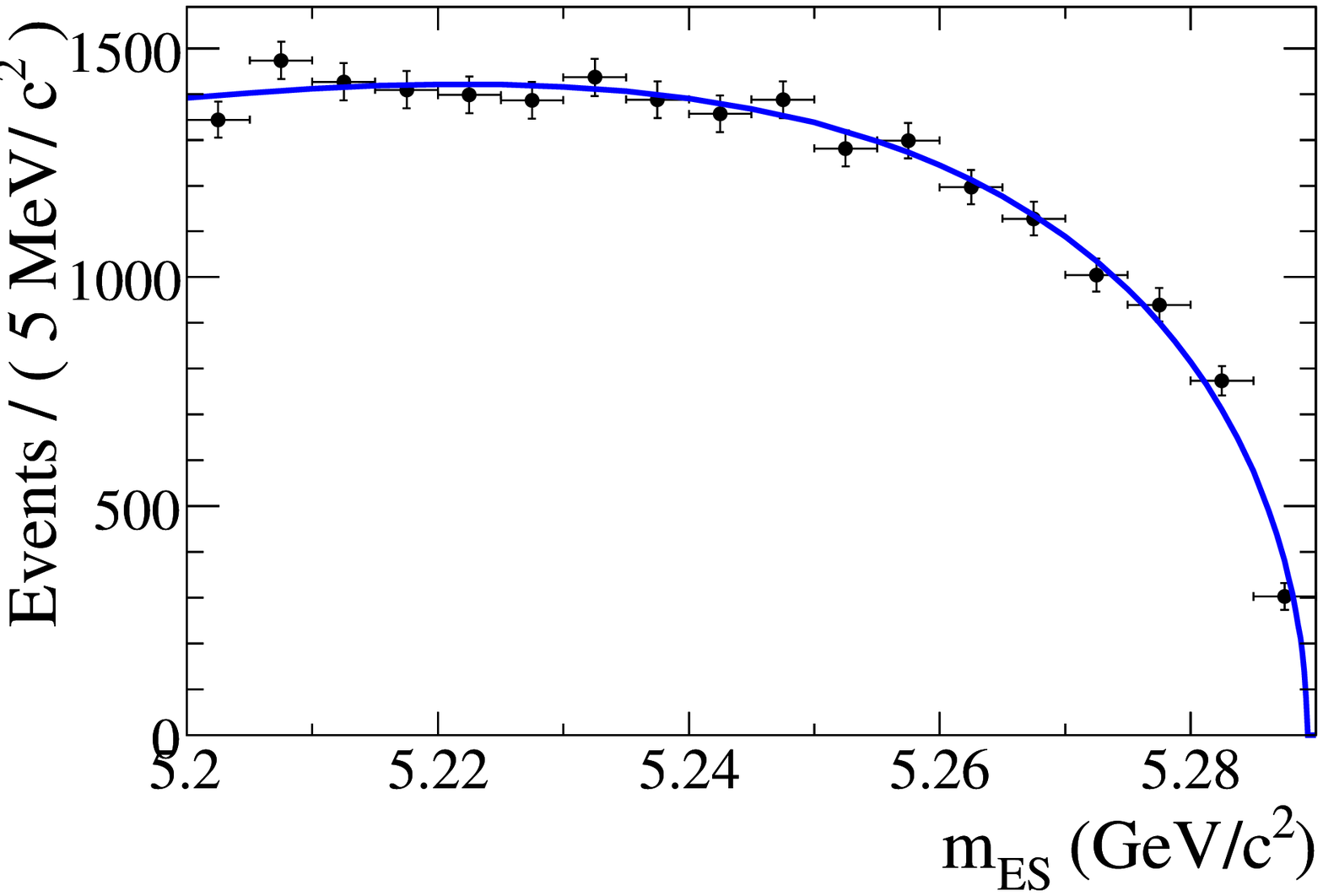}
\includegraphics[width=0.7\linewidth]{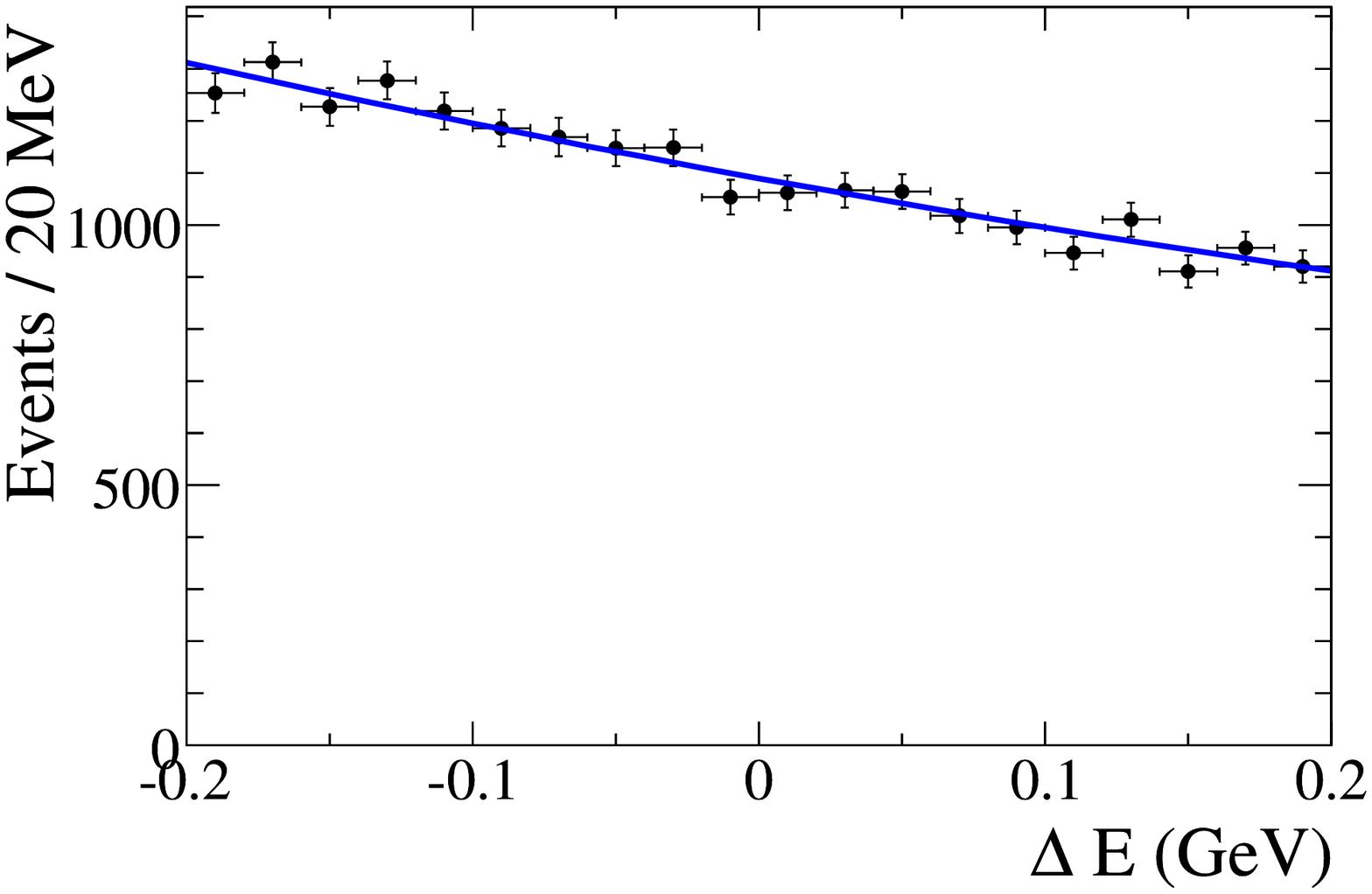}
\hspace*{2.0mm}\includegraphics[width=0.78\linewidth]{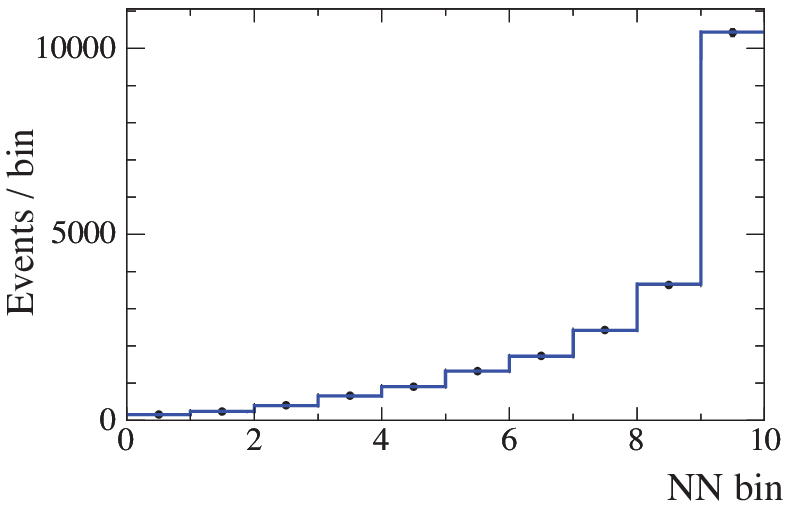}
\caption{ \Bztopizpiz background plots with signal 
subtracted using the \sPlots technique.
From top to bottom: \mes, \de, and \emph{NN}. The points with error bars
show the data, and the line
in each plot shows the corresponding PDF.
}
\label{fig:pizpizBkg}
\end{center}
\end{figure}

Results from the ML fit for the \Bztopizpiz decay mode are
summarized in Table~\ref{tab:resultsA}. 
\sPlots  of \mes, \de, and \emph{NN} for \Bztopizpiz
are shown in Fig.~\ref{fig:pizpiz}, and for the \qq
background in Fig.~\ref{fig:pizpizBkg}.

\begin{table*}[!htbp]
\caption{Results for the \Bztopizpiz\ and \Bztokspiz\ decay modes,
  showing the signal yield $N_{\rm sig}$, efficiency, branching
  fraction, and \CP-violation parameter $C$ for each mode.  When two
  uncertainties are given, the first is statistical and the second is
  systematic.  Uncertainties for the signal yields are statistical,
  and those for the efficiencies are systematic.  }
\label{tab:resultsA}
\begin{center}
\begin{tabular}{ccccc}
\hline\hline
 & $N_{\rm sig}$  & Efficiency ($\%$)      & Branching fraction ($10^{-6}$)  & $C$         \\ \hline 
\Bztopizpiz & $247\pm 29$  & $28.8\pm1.8$ & $ 1.83 \pm 0.21 \pm 0.13 $ & $-0.43 \pm 0.26 \pm 0.05$  \\
\Bztokspiz & $556 \pm 32$  & $34.2\pm1.2$ & 
        $5.1 \pm 0.3 \pm 0.2$ &  
         \\
\hline\hline
\end{tabular}
\end{center}
\end{table*}

The various systematic uncertainties for the \Bztopizpiz decay mode are listed in 
Tables~\ref{tab:pizpizsyst-fit} and~\ref{tab:pizpizsyst-br}.
The uncertainty in the efficiency is
dominated by a $3\%$ systematic uncertainty per \piz, estimated from a
study of $\tau \to \pi\piz\nu_{\tau}$ decays. An uncertainty of 
$1.0\%$ is due to the resolution of the signal shape, and an additional
uncertainty of $0.5\%$ is due to the limited knowledge of the \mes and \de
peak positions in data. These are estimated by shifting the \mes and \de means
and resolutions by amounts determined from MC--data comparison in a
control sample of $\Bp\to \pi^+\pi^0$ events. An
uncertainty of 1.5\%, determined from the \Bflav\ sample, is due to the
$\cossph$ requirement. A 1.1\% uncertainty is assigned to the 
number of \BB\ events in the data sample.
Systematic uncertainties involving the ML
fit are evaluated by varying the PDF parameters and refitting the
data. These contribute an uncertainty of 8.3 events to the
branching-fraction measurement and an uncertainty of 0.055 to
$\cpizpiz$.

\begin{table}[!htbp]
\caption{ 
  Systematic uncertainties on the \Bztopizpiz signal yield $N_{\piz\piz}$
  and direct \CP asymmetry $\cpizpiz$.
  The total uncertainty is the sum in quadrature of the individual uncertainties.}
\label{tab:pizpizsyst-fit}
\smallskip
\begin{center}
\begin{tabular}{l|c|c}
\hline \hline
Source                &   $N_{\piz\piz}$   & $\cpizpiz$ \\ 
\hline
Peaking background    &   $ 4.9$           & $ 0.030$     \\
Tagging               &   $ 0.35$          & $ 0.034$     \\
Background shape      &   $ 5.5$           & $ 0.023$     \\
Signal shape          &   $ 3.8$           & $ 0.020$     \\ 
\hline
Total fit systematic uncertainty &   $ 8.3$  &  $ 0.055$     \\ 
\hline 
\hline
\end{tabular}
\end{center}
\end{table}

\begin{table}[!htbp]
\caption{ 
  Relative systematic uncertainties on the \Bztopizpiz branching fraction. 
  The total uncertainty is the sum in quadrature of the relative uncertainties on
  the signal yield (from Table~\ref{tab:pizpizsyst-fit}), the signal efficiency, and the number of $\BB$ pairs. }
\label{tab:pizpizsyst-br}
\smallskip
\begin{center}
\begin{tabular}{l|c}
\hline \hline
Source                 & ${\cal B}(\Bztopizpiz)$  \\ 
\hline
Signal yield syst. uncertainty &  $3.4\%$    \\ 
\hline
\piz efficiency        & $6.0\%$ \\
$\cossph$ selection    & $1.5\%$ \\
neutrals resolution    & $1.0\%$ \\
\mes and \de shape     & $0.5\%$ \\ 
\hline
Number of $\BB$ pairs          & $ 1.1\%$  \\
\hline
Total systematic uncertainty & $7.2\%$ \\ 
\hline 
\hline
\end{tabular}
\end{center}
\end{table}


\subsection{\boldmath \Bztokspiz\ Results}
\label{sec:kspizResults}
The efficiency and branching fraction measured for the \Bztokspiz\ decay 
mode are summarized in Table~\ref{tab:resultsA} 
(CP-violation parameters have been reported in
Ref.~\cite{Aubert:2008ad}).
 
\begin{figure}[!htbp]
\begin{center}
\includegraphics[width=1.0\linewidth]{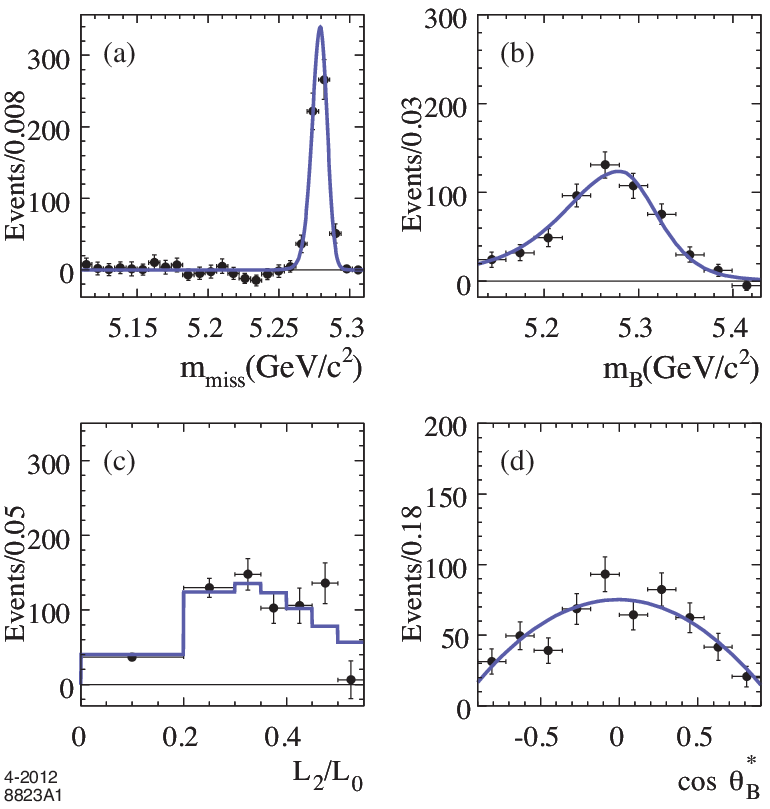}
\includegraphics[width=0.6\linewidth]{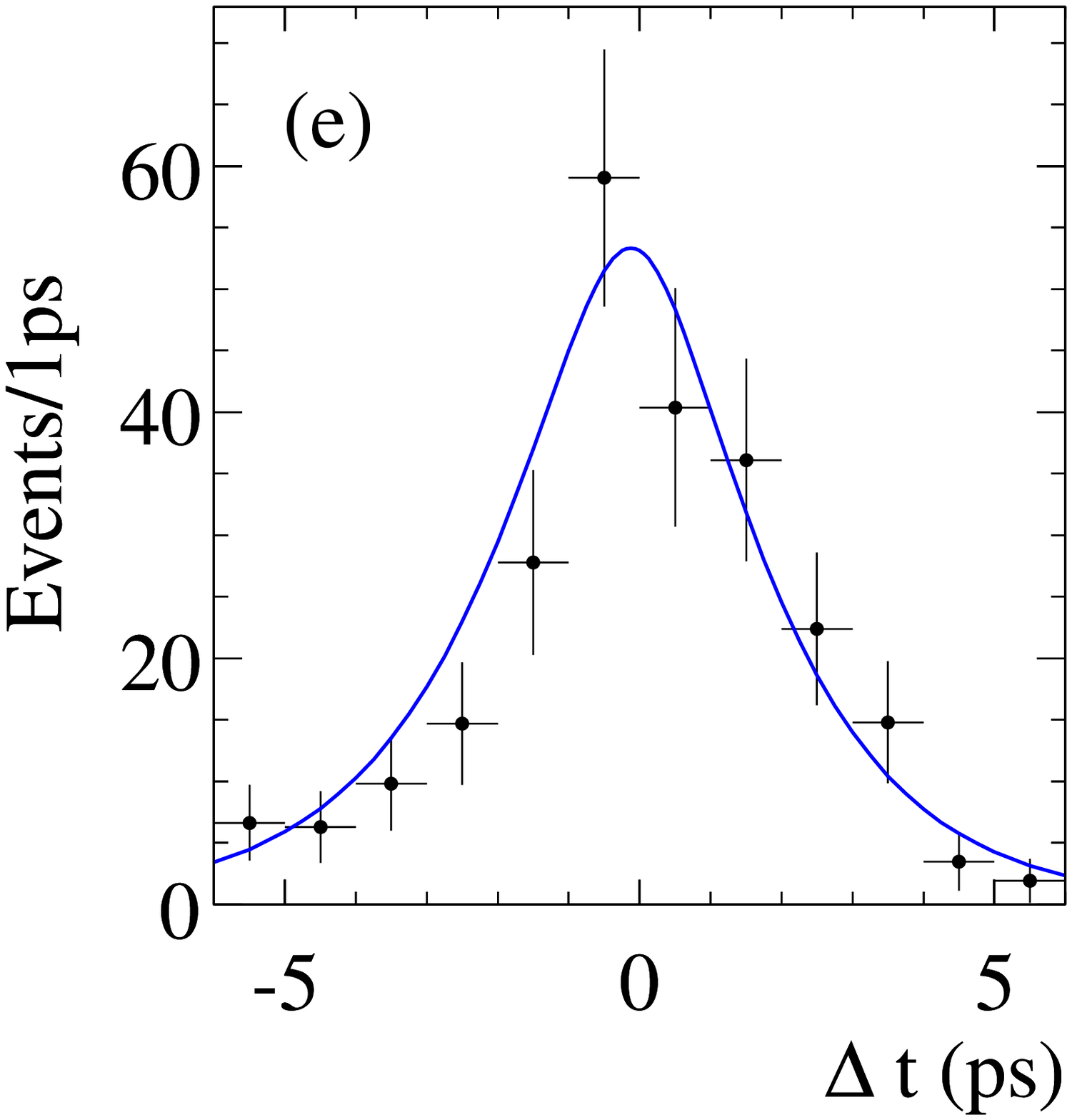}
\caption{ \sPlots of the  (a) $\mmiss$, (b) $\mb$, (c) $L_2/L_0$,
(d)~$\costhetacms$, and (e) $\deltat$ distributions for signal  events
in the \Bztokspiz\ sample.  
The points with error bars represent
the data, and the lines show the shapes of signal PDFs as
obtained from the ML fit. 
}
\label{fig:kspi0_sig}
\end{center}
\end{figure}

\begin{figure}[!htbp]
\begin{center}
\includegraphics[width=1.0\linewidth]{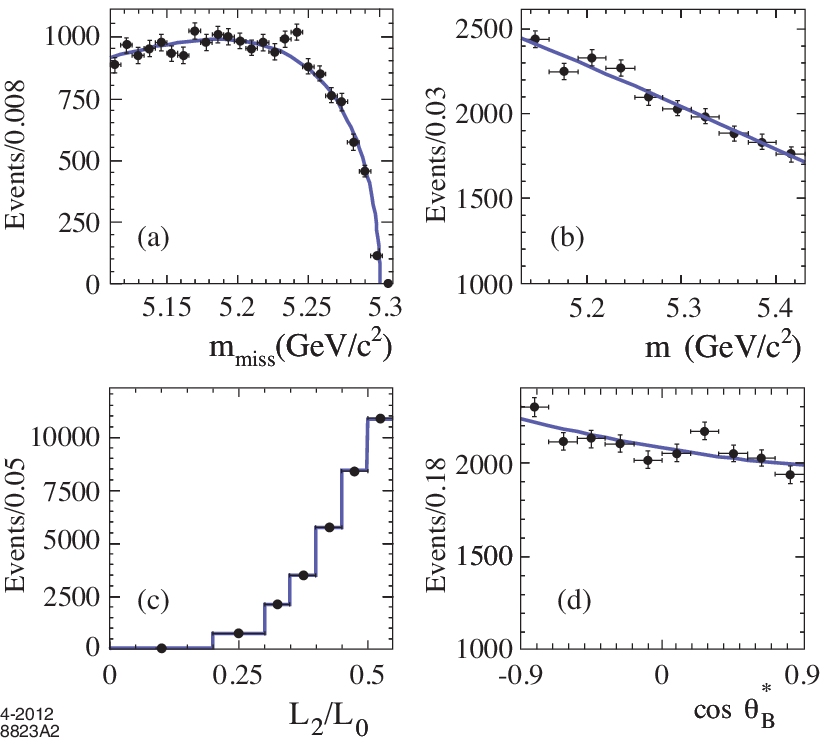}
\includegraphics[width=0.6\linewidth]{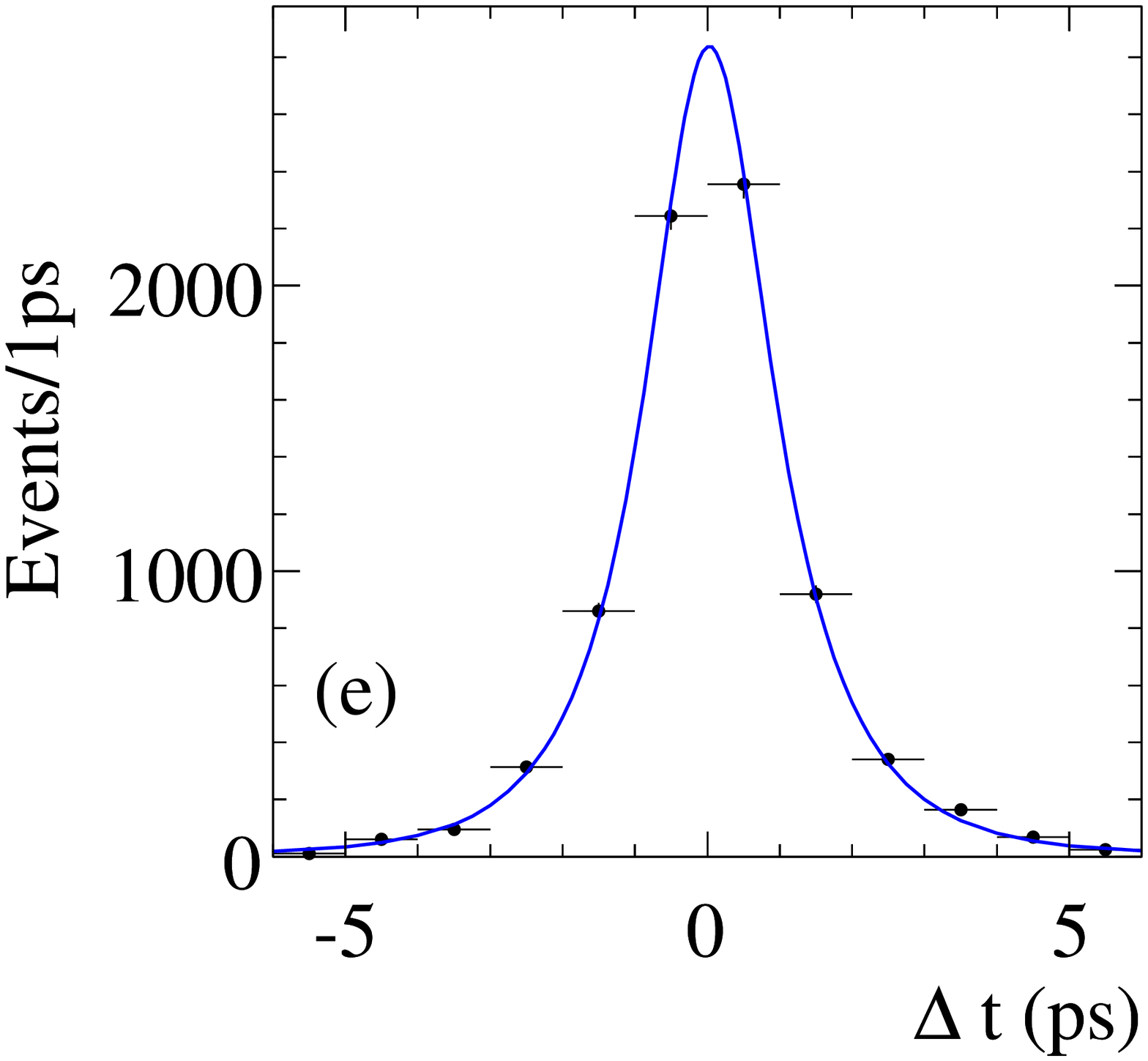}
\caption{ \sPlots of the  (a) $\mmiss$, (b) $\mb$, (c) $L_2/L_0$,
(d)~$\costhetacms$, and (e) $\deltat$ distributions for background  events
in the \Bztokspiz\ sample.  
The points with error bars represent
the data, and the lines show the shapes of signal PDFs as
obtained from the ML fit. 
}
\label{fig:kspi0_bkg}
\end{center}
\end{figure}

We show \sPlots of $m_{\rm miss}$, $m_B$, 
$L_2/L_0$, and $\costhetacms$ for signal events 
in Fig.~\ref{fig:kspi0_sig} and for background events in 
Fig.~\ref{fig:kspi0_bkg}.

The systematic uncertainties on the branching fraction ${\cal B}(\Bztokspiz)$
are summarized in Table~\ref{tab:BFsys}.
The uncertainty on the efficiency of the \Ks\ reconstruction is obtained
from detailed comparison of inclusive \Ks\ candidates in data and MC. 
The $\piz$ efficiency uncertainty is evaluated from the ratio of 
branching fractions $\BR(D^0\to K^-\pi^+\piz) / \BR(D^0\to K^-\pi^+)$.
To compute the systematic uncertainty associated with the statistical
precision on the parameters of the likelihood function, we shift
each parameter by its associated uncertainty and repeat the fit.
For \deltat and the tagging parameters, the uncertainty is obtained
from the fit to the \Bflav\ sample, while for the other parameters it
is obtained from MC. This uncertainty
accounts for the size of the sample used for determining the shape
of the likelihood function in Eq.~(\ref{eq:ml}).  A systematic uncertainty
associated with the data--MC agreement in the shape of the
signal PDFs is evaluated by taking the largest deviation observed when the
parameters of the individual signal PDFs for \mmiss, \mb, $L_2/L_0$,
and \costhetacms{} are allowed to vary in the fit. The output values of the PDF
parameters are also used to assign a systematic uncertainty to the 
efficiency of the event selection requirements on the likelihood variables, by comparing the
efficiency in data to that in the MC. 
We evaluate the
systematic uncertainty due to the neglected correlations among fit
variables using a set of MC experiments, in which we embed signal
events from a full detector simulation with events generated from the
background PDFs. Since the shifts are small and only marginally
significant, we use the average relative shift in the yield as
the associated systematic uncertainty.

In the fit we neglect background from $B$ decays, which is estimated
from simulation to contribute of order 0.1\% of the total background.
To account for a bias due to this, we study in
detail the effect of a number of specific $B$ decay channels that
dominate this type of background, notably $B^+\ra\rho^+\KS$, $B^+\ra
K^{*+}\piz$, and $B^+\ra\KS\piz\pi^+$.
We embed these simulated $B$-background
events in the data set and find the average shift in the fit
signal yield to be $+5.2$ events. We adjust the signal yield
accordingly and use half of the bias as a systematic uncertainty.

For the branching fraction, additional systematic uncertainties originate 
from the 
uncertainty on the 
selection efficiency, 
the number of \BB{} pairs in the data sample
(1.1\%),
and the branching fractions 
${\cal B}(K^0_S \to \pi^+\pi^-)$ 
and ${\cal B}(\pi^0 \to \gamma \gamma)$~\cite{pdg}. 

\begin{table}[!htbp]
    \caption{ Summary of dominant contributions to the systematic uncertainty
    on the measurement of ${\cal B}(\Bztokspiz)$ \label{tab:BFsys}}
  \begin{center}
    \begin{tabular}{l|c}
      \hline\hline           
  Source                                               & $\sigma({\cal B}(\Bztokspiz))$ (\%) \\
      \hline  
     $\pi^0$ efficiency                         & $3.0$ \\
     $\KS$ efficiency                           & $0.5$ \\
     Selection criteria                         & $1.5$ \\
    PDF-parameters precision                    & $0.22$ \\
    Shape of signal PDFs                        & $0.45$ \\
    $\BB$ background                            & $0.47$  \\
    Correlations                                & $0.40$  \\
    Resolution function                         & $0.49$  \\
    Number of $\BB$ pairs                            & $1.1$  \\
      \hline
      Total                                         & $3.7$  \\
      \hline\hline
    \end{tabular}
  \end{center}
\end{table}

\section{RESULTS FOR \boldmath $\delalph$ and $\alpha$}
\label{sec:alpha}

\begin{figure}[!htbp]
\begin{center}
\includegraphics[width=0.69\linewidth]{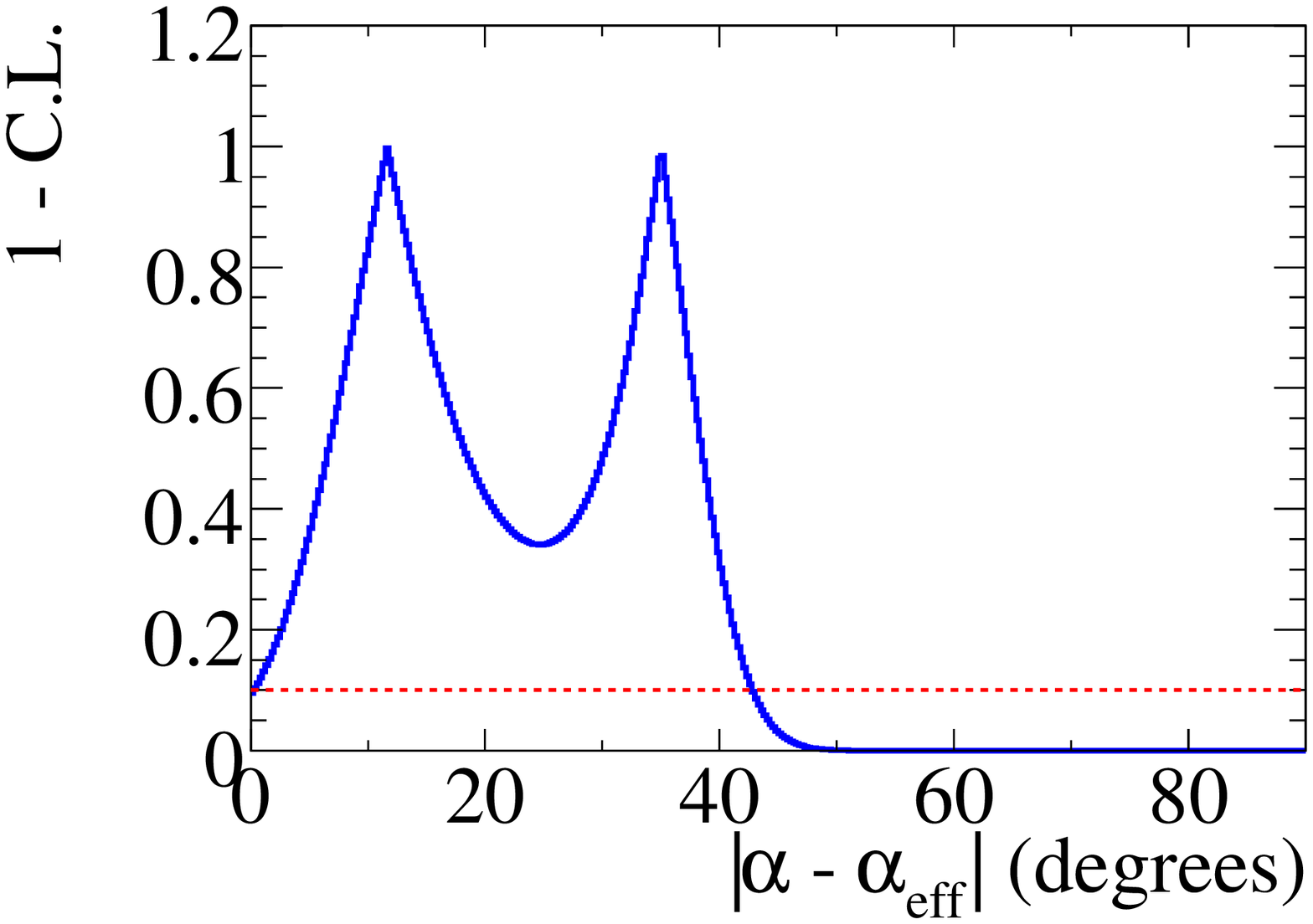}
\includegraphics[width=0.69\linewidth]{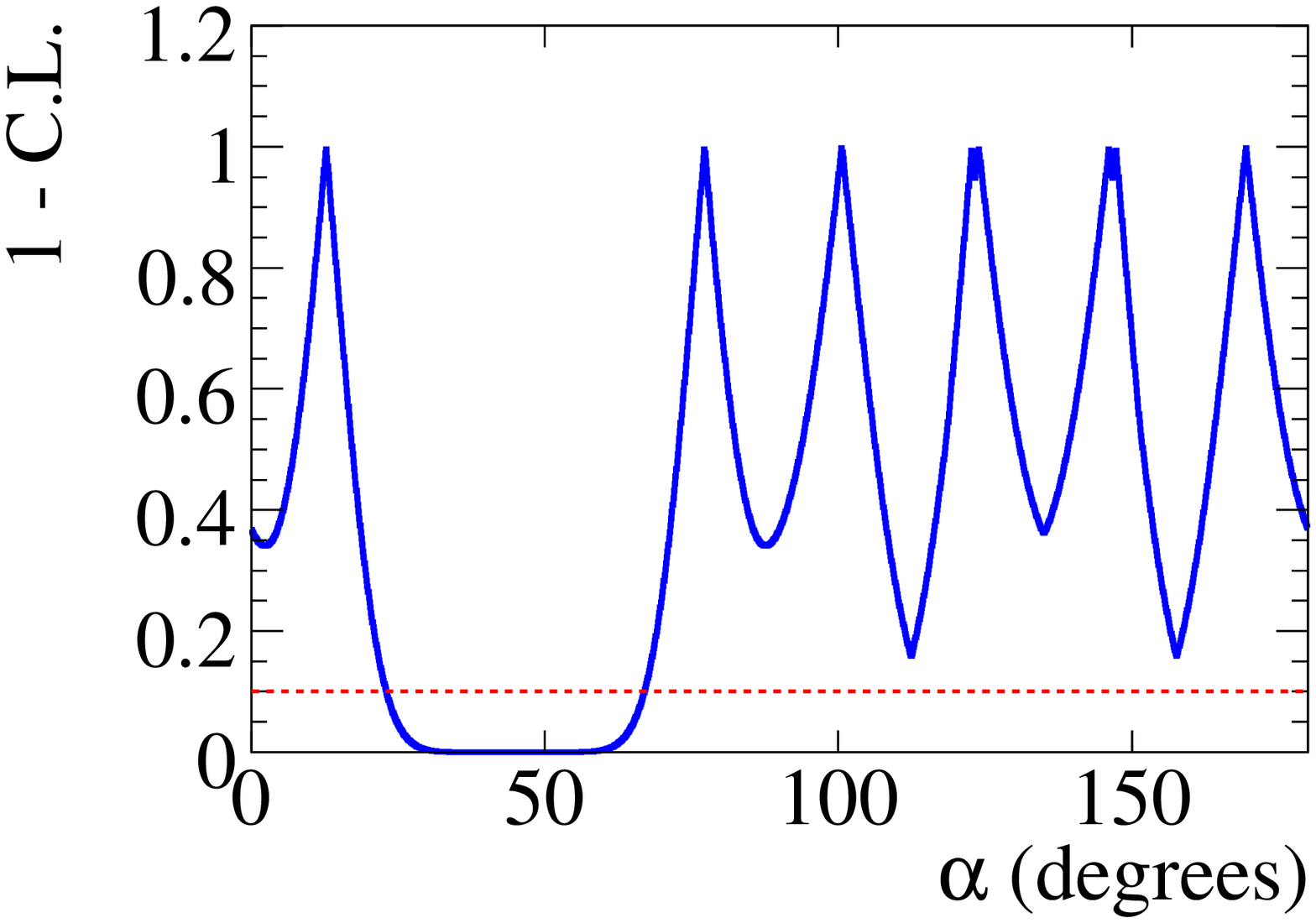}
\end{center}
\vspace{0.1cm}
\caption{
 (Top)
  Constraint on  $\delalph = \alpha - \alphaeff$,
  expressed as one minus the confidence level as a function of
  $|\delalph|$.  We find an upper bound on $\left|\delalph\right|$ of
  $43^\circ$ at the 90\% C.L. 
(Bottom)
 constraint on the CKM angle $\alpha$.
  We exclude the range $[23^\circ, 67^\circ]$ in $\alpha$ at the 90\% C.L.
Only the isospin-triangle relations and the expressions in
  Eq.~(\ref{eq:asymmetry}) are used in this constraint. 
}
\label{fig:alpha}
\end{figure}

We combine our results for \BR(\Bztopizpiz) with the branching fractions
$\BR(\Bztopippim) = (5.5 \pm 0.4 \pm 0.3) \times 10^{-6}$ 
and $\BR(B^{\pm}\to\pi^{\pm}\pi^0) = (5.02 \pm 0.46 \pm 0.29)\times 10^{-6}$ 
previously measured by \babar~\cite{BabarBRPiPi,pi0pi0_BaBar} to evaluate the
constraints on both the penguin contribution to $\alpha$ and on the
CKM angle $\alpha$ itself. Constraints are evaluated by scanning the
parameters  $\left|\delalph\right|$ and $\alpha$, and
then calculating the $\chi^{2}$ for the five amplitudes ($A^{+0}$,
$A^{+-}$, $A^{00}$, $\Abar^{+-}$, $\Abar^{00}$) from our
measurements and the isospin-triangle relations~\cite{ref:CKMfitter}. Each $\chi^2$ value is
converted to a confidence level, shown in Fig.~\ref{fig:alpha}
for $\delalph$ and $\alpha$. 
The $\alpha$ plot exhibits six clear peaks, a result of the 
eight-fold trigonometric
ambiguity in the extraction of $\alpha$ and the fact that 
two pairs of peaks are nearly merged.
The upper bound on  $\left|\delalph\right|$ is $43^\circ$ at the 90\% C.L.,
and the range $[23^\circ, 67^\circ]$ in $\alpha$ 
is excluded at the 90\% C.L.
  The point $\alpha=0$, which corresponds to no \CP violation, and
  the values of $\alpha$ near 0 or $\pi$ can be excluded with additional
  physics input~\cite{pi0pi0_BaBar,UTFit_2007_pipi}.
If we consider only the solution preferred in the
SM~\cite{Gronau2007}, $\alpha$ lies in the range $[71^\circ,109^\circ]$ at the 68\% C.L.  This is consistent with the more 
restrictive constraints on $\alpha$
obtained from analysis of the $B\to\rho\rho$
system~\cite{ref:rhorho}, as well as those 
from $\Bz\to(\rho\pi)^0$~\cite{ref:rhopi} and $\Bz\to a_1 \pi$~\cite{ref:a1pi}.

\section{CONCLUSIONS}
\label{sec:Conclusions}

We measure the \CP-asymmetry parameters 
\begin{align*}
   \spipi & =   -0.68 \pm 0.10 \pm 0.03, \\
   \cpipi & =   -0.25 \pm 0.08 \pm 0.02, \\
   \akpi & = -0.107 \pm 0.016 ^{+0.006}_{-0.004}, \\
   \cpizpiz & =  -0.43 \pm 0.26 \pm 0.05
\end{align*}
and CP-averaged branching fractions
\begin{align*}
   \BR(\Bztopizpiz) & = ( 1.83 \pm 0.21 \pm 0.13 ) \times 10^{-6}, \\
   \BR(\Bztokzpiz) & = ( 10.1 \pm 0.6 \pm 0.4 ) \times 10^{-6}.           
\end{align*}
We find a 68\% C.L. region for $\alpha$ of $[71^\circ,109^\circ]$ and exclude
values in the range $[23^\circ, 67^\circ]$ at the 90\% C.L.
We observe direct \CP
violation in $\Bz\to\Kp\pim$ with a significance of $6.1\sigma$ and in
\Bztopipi with a significance of $6.7\sigma$, including systematic uncertainties.  Ignoring
color-suppressed tree amplitudes, the charge asymmetries in $\Kp\pim$
and $\Kp\piz$ should be equal~\cite{ref:SumRule1}, which is not
supported by recent \babar\ and Belle
data~\cite{BaBarPRL2007,pi0pi0_BaBar,BelleNature2008}.  These results
might indicate a large color-suppressed amplitude, an enhanced
electroweak penguin, or possibly new-physics effects~\cite{ref:NP}.

Our result for $\BR(\Bztokzpiz)$ is consistent with the 
sum-rule prediction~\cite{ref:SumRule1, ref:SumRule2}
$
\BR(K^0\piz)^{\rm sr} =  \frac12
   \left(\BR(K^+\pim) + 
       \frac{\tau_0}{\tau_+} \left[\BR(K^0\pip)- 2 \BR(K^+\piz)]\right]
   \right)   
   = (8.4 \pm 0.8)\times 10^{-6},
$
obtained using the currently published
results~\cite{pi0pi0_BaBar,BabarBRPiPi,ref:BaBarK0K0,ref:BelleKpiData,ref:CLEOdata}
for the three $B\ra K\pi$ rates on the right-hand side of this
equation and the lifetimes $\tau_+$ and $\tau_0$ of the charged 
and neutral $B$ mesons.

The results presented here supersede those of our prior
publications~\cite{BaBarPRL2007,pi0pi0_BaBar,ref:BaBarK0pi0}.

\section{ACKNOWLEDGMENTS}
\label{sec:Acknowledgments}

\input acknowledgements




\end{document}

%% file: authors_feb2012.tex
%
\author{J.~P.~Lees}
\author{V.~Poireau}
\author{V.~Tisserand}
\affiliation{Laboratoire d'Annecy-le-Vieux de Physique des Particules (LAPP), Universit\'e de Savoie, CNRS/IN2P3,  F-74941 Annecy-Le-Vieux, France}
\author{J.~Garra~Tico}
\author{E.~Grauges}
\affiliation{Universitat de Barcelona, Facultat de Fisica, Departament ECM, E-08028 Barcelona, Spain }
\author{A.~Palano$^{ab}$ }
\affiliation{INFN Sezione di Bari$^{a}$; Dipartimento di Fisica, Universit\`a di Bari$^{b}$, I-70126 Bari, Italy }
\author{G.~Eigen}
\author{B.~Stugu}
\affiliation{University of Bergen, Institute of Physics, N-5007 Bergen, Norway }
\author{D.~N.~Brown}
\author{L.~T.~Kerth}
\author{Yu.~G.~Kolomensky}
\author{G.~Lynch}
\affiliation{Lawrence Berkeley National Laboratory and University of California, Berkeley, California 94720, USA }
% ...
\collaboration{The \babar\ Collaboration}
\noaffiliation

%% file: acknowledgements.tex
We are grateful for the 
extraordinary contributions of our \pep2\ colleagues in
achieving the excellent luminosity and machine conditions
that have made this work possible.
The success of this project also relies critically on the 
expertise and dedication of the computing organizations that 
support \babar.
The collaborating institutions wish to thank 
SLAC for its support and the kind hospitality extended to them. 
This work is supported by the
US Department of Energy
and National Science Foundation, the
Natural Sciences and Engineering Research Council (Canada),
the Commissariat \`a l'Energie Atomique and
Institut National de Physique Nucl\'eaire et de Physique des Particules
(France), the
Bundesministerium f\"ur Bildung und Forschung and
Deutsche Forschungsgemeinschaft
(Germany), the
Istituto Nazionale di Fisica Nucleare (Italy),
the Foundation for Fundamental Research on Matter (The Netherlands),
the Research Council of Norway, the
Ministry of Education and Science of the Russian Federation, 
Ministerio de Ciencia e Innovaci\'on (Spain), and the
Science and Technology Facilities Council (United Kingdom).
Individuals have received support from 
the Marie-Curie IEF program (European Union) and the A. P. Sloan Foundation (USA).

%% file: paper.bbl
\begin{thebibliography}{99}
\bibitem{refConj}
  The use of charge-conjugate modes is implied throughout this paper unless otherwise noted.
\bibitem{rho-a1} We use the shorthand notation $\rho$ and $a_1$ to
refer to the $\rho(770)$ and $a_1(1260)$, respectively.
\end{thebibliography}
